\documentclass[a4paper, 12pt, oneside]{article}
\usepackage[cp1251]{inputenc} 
\usepackage[english, russian]{babel} 
\usepackage{ graphics,amsfonts, amsmath,bm,amssymb, array, eucal}
\usepackage[dvips]{graphicx}
\usepackage[flushleft,small]{caption2} 



\begin{document}
\newcommand{\mc}[1]{\mathcal{#1}}
\newcommand{\E}{\mc{E}}
\thispagestyle{empty}
\large

\renewcommand{\abstractname}{\,Abstract}
\renewcommand{\refname}{\begin{center} REFERENCES\end{center}}

 \begin{center}
\bf Longitudinal electric conductivity and dielectric permeability
in quantum plasma with constant collision frequency  in Mermin' approach
\end{center}\medskip
\begin{center}
  \bf A. V. Latyshev\footnote{$avlatyshev@mail.ru$} and
  A. A. Yushkanov\footnote{$yushkanov@inbox.ru$}
\end{center}\medskip

\begin{center}
{\it Faculty of Physics and Mathematics,\\ Moscow State Regional
University, 105005,\\ Moscow, Radio str., 10--A}
\end{center}\medskip

\begin{abstract}
Detailed deducing of formulas
for longitudinal electric conductivity and dielectric permeability
in the quantum degenerate collisional plasma with constant collision frequency
in Mermin' approach is given.
The kinetic Schr\"{o}din\-ger---Boltzmann equation in momentum space
in relaxation approximation is used.

It is shown that when  collision frequency of  plasma particles
tends  to zero (plasma passes to collisionless one), the deduced
formula for dielectric function passes to the known Lindhard' formula
for collisionless plasmas.

It is shown that the deduced formula for dielectric permeability
coincides with known Mermin's formula.

Graphic research of the real and imaginary parts
of dielectric function is made.

Graphic comparison of the real and imaginary parts
of dielectric function  for quantum and classical plasma also is made.

The module of derivative dielectric function also
has been investigated graphically.


\end{abstract}

\begin{center}
{\bf Introduction}
\end{center}

In the known work of Mermin \cite{Mermin} on the basis of the
analysis of nonequilib\-rium density matrix in $\tau $--approximation
has been obtained expression for longitu\-di\-nal dielectric permeability of
quantum collisional plasmas for case of constant collision frequency of plasmas
particles.

Earlier in the work of Klimontovich and Silin \cite{Klim} and after that in
the work of Lindhard \cite{Lin} has been obtained
expression for longitudinal and transverse dielectric permeability of quantum
collisionless plasmas.

By Kliewer and Fuchs \cite{Kliewer} it has been shown, that
direct generalisation of formulas of Lindhard on the case of collisional plasmas
is incorrectly. This lack for the longitudinal dielectric
permeability has been eliminated in the work of Mermin \cite{Mermin}.
Next in the works \cite{Das} and \cite{Persson} has been given attempt
to deduce Mermin's formula.

For collisional plasmas correct formulas longitudinal and transverse
electric conductivity and dielectric permeability are received
accordingly in works \cite{Long} and \cite{Trans}. In these works
Wigner---Vlasov---Boltzmann kinetic equation
in relaxation approximation in coordinate space is used. In work
\cite{Trans2} the formula for transverse electric conductivity has been deduced
for quantum collisional plasmas with use of the kinetic equation
by Mermin' approach (in momentum space).

In the present article formulas for longitudinal electric conductivity
and dielectric permeability in the quantum degenerate collisional
plasma with constant collisional frequency by Mermin' approach are received.
The kinetic Schr\"{o}din\-ger---Boltzmann equation in momen\-tum space in relaxation approximation is
used.

It is shown also, that when frequency of collisions of particles of plasma
tends  to zero (plasma passes to collisionless one), the deduced formula
for dielectric function passes to the known Lindhard' formula for collisionless
plasmas. It is shown, that when frequency of collisions is a constant,
the deduced formula for dielectric function
passes in known Mermin' formula.

It is shown that the deduced formula for dielectric permeability
coin\-ci\-des with known Mermin's formula.

Last years there is a steady interest to research
properties of quantum plasma \cite{Manf}--\cite{Ropke}.

\begin{center}
  \bf 1. Kinetic Schr\"{o}dinger---Boltzmann equation for density matrix
\end{center}

Let the vector potential of an electromagnetic field is
harmonious, i.e. changes as
$
{\varphi}({\bf r},t)={\varphi}({\bf r})\exp(-i \omega t).
$
Relation between scalar potential and intensity of the electric
field it is given by expression
$
{\bf E}({\bf q})=-\nabla{\varphi}({\bf q}).
$
The equilibrium matrix of density looks like
$$
{\tilde \rho}=\dfrac{1}{\exp\dfrac{H-\mu}{k_BT}+1},\qquad \mu=\mu_0+\delta\mu.
$$

Here $T $ is the temperature, $k_B$ is the Boltzmann constant,
$\mu_0$ is the chemical potential of plasma in an equilibrium condition,
$\delta\mu $ is the correction to the chemical potential, caused
presence of variable electric field, $H $ is the Hamiltonian.

Hamiltonian looks like here $H=H_0+H_1$, where
$H_0={\bf p}^2/2m,\; H_1=e\varphi$.
Here $m, e$ are mass and charge of electron, ${\bf p}=\hbar{\bf k}$
is the electron momentum.

Let's designate an equilibrium matrix of density in absence of an external field
through ${\tilde \rho}_0$:
$$
{\tilde \rho}_0=\dfrac{1}{\exp \dfrac{H_0-\mu_0}{k_BT}+1}.
$$

Density matrix it is possible to present an equilibrium matrix of density
in the form
$$
{\tilde \rho}={\tilde \rho}_0+{\tilde \rho}_1.
$$

Here ${\tilde \rho}_1$ is the correction to the equilibrium matrix of density,
caused by presence of an electromagnetic field.

In linear approximation we receive
$$
[H, {\tilde \rho}\,]=[H_0, {\tilde \rho}_1]+[H_1,{\tilde\rho}_0],
$$
and
$$
[H, {\tilde \rho}\,]=0.
$$

Here $[H, {\tilde \rho}\,]=H {\tilde \rho}-{\tilde \rho}H$ is
the commutator.

Let's notice that the vector $|{\bf k}\rangle $ is the eigen vector
of operators $H $ and $ \bf p $. Thus
$$
H|{\bf k}\rangle=\E_{\bf k}|{\bf k}\rangle,\quad
\langle{\bf k}|H=\E_{\bf k}\langle{\bf k}|, \quad
{\bf p}|{\bf k}\rangle=\hbar{\bf k}|{\bf k}\rangle, \quad
\langle{\bf k}|{\bf p}=\hbar{\bf k}\langle{\bf k}|.
$$

Let's notice that for the operator $L $ the relationship is carried out
$$
\langle{\bf k}_1|L|{\bf k}_2\rangle=
\dfrac{1}{(2\pi)^3}\int\exp(-i{\bf k}_1{\bf r})L\exp(i{\bf k}_2{\bf
r})d{\bf r}.
$$

By means of this relation it is received
$$
\left\langle{\bf k}_1\left|[H_0, {\tilde \rho}_{1}]\right|{\bf k}_2
\right\rangle=-\left\langle{\bf k}_1\left|[H_1, {\tilde \rho}_{0}]\right|
{\bf k}_2\right\rangle.
$$

Let's write down this equality in the explicit form
$$
\left\langle{\bf k}_1\left|H_0{\tilde \rho}_1\right|{\bf k}_2
\right\rangle - \left\langle{\bf k}_1\left|{\tilde \rho}_1 H_0\right|{\bf k}_2
\right\rangle=-\left\langle{\bf k}_1\left|H_1 {\tilde \rho}_0\right|
{\bf k}_2\right\rangle+\left\langle{\bf k}_1\left|{\tilde \rho}_0 H_1 \right|
{\bf k}_2\right\rangle.
$$

From here we receive that
$$
(\E_{{\bf k}_1}-\E_{{\bf k}_2})\langle {\bf k}_1|{\tilde \rho}_1|{\bf k}_2)
\rangle=
(f_{{\bf k}_1}-f_{{\bf k}_2})\langle{\bf k}_1|H_1|{\bf k}_2\rangle=
$$
$$
=e(f_{{\bf k}_1}-f_{{\bf k}_2})\langle{\bf k}_1|\varphi|{\bf k}_2\rangle.
$$
Here
$$
\E_{{\bf k}}=\dfrac{\hbar^2 {\bf k}^2}{2m},\qquad
f_{{\bf k}}=\dfrac{1}{\exp\dfrac{E_{{\bf k}}-\mu_0}{k_BT}+1}.
$$

Considering that
$$
\langle{\bf k}_1|\varphi|{\bf k}_2\rangle=\dfrac{1}{(2\pi)^3}
\int\exp(-i({\bf k}_1-{\bf k}_2){\bf r})
\varphi({\bf r})d{\bf r}=\varphi({\bf k}_1-{\bf k}_2),
$$
we receive
$$
(\E_{{\bf k}_1}-\E_{{\bf k}_2}){\tilde \rho}_{1}({\bf k}_1,{\bf k}_2)=
e(f_{{\bf k}_1}-f_{{\bf k}_2})\varphi({\bf k}_1-{\bf k}_2).
$$

The kinetic Schr\"{o}dinger---Boltzmann equation for the density matrix
in $\tau$--approximation with
constant frequency of collisions looks like
$$
i\hbar \dfrac{\partial \rho}{\partial t}=[H,\rho]+\dfrac{i\hbar}{\tau}({\tilde
\rho}-\rho),
$$
or
$$
\dfrac{\partial \rho}{\partial t}=-\dfrac{i}{\hbar}[H,\rho]+\nu({\tilde
\rho}-\rho).
\eqno{(1.1)}
$$

Here $\tau$ is the average time of free electrons path,
$ \nu=1/\tau $ is the frequency of collisions.

In linear approximation  the density matrix we will search in the form
$$
\rho={\tilde \rho}_0+{\rho}_1.
$$

Then in linear approach the equation (1.2) looks like
$$
i\hbar \dfrac{\partial \rho_1}{\partial t}=[H_0, \rho_1]+[H_1,\rho_0]+
i\nu\hbar (\tilde{\rho}_1-\rho_1).
\eqno{(1.2)}
$$

Let's consider, that $ \rho_{1}\sim \exp(-i\omega t) $.
From here for $ {\rho}_{1} $ we receive the relation
$$
\hbar \omega\langle{\bf k}_1|\rho_1|{\bf k}_2\rangle=
(\E_{{\bf k}_1}-\E_{{\bf k}_2})
\langle{\bf k}_1|\rho_1|{\bf k}_2\rangle
-e(f_{{\bf k}_1}-f_{{\bf k}_2}){\varphi}({\bf k}_1-{\bf k}_2)+
$$
$$
+i\hbar\nu\langle{\bf k}_1|{\tilde \rho}_1-\rho_1|{\bf k}_2\rangle.
\eqno{(1.2)}
$$

We transform this equation to the form
$$
(\E_{{\bf k}_2}-\E_{{\bf k}_1}+\hbar \omega+i \nu \hbar)
\langle {\bf k}_1|\rho_1|{\bf k}_2\rangle=-
e(f_{{\bf k}_1}-f_{{\bf k}_2})\langle {\bf k}_1|\varphi|{\bf k}_2\rangle+$$$$+
i \nu \hbar \langle {\bf k}_1|\tilde\rho_1|{\bf k}_2\rangle.
$$

From this equation we obtain
$$
\langle {\bf k}_1|\rho_1|{\bf k}_2\rangle=\rho_1({\bf k}_1,{\bf k}_2)=
-e\dfrac{f_{{\bf k}_1}-f_{{\bf k}_2}}
{\E_{{\bf k}_2}-\E_{{\bf k}_1}+\hbar \omega+i \nu \hbar}
\varphi({\bf k}_1-{\bf k}_2)+
$$
$$
+\dfrac{i \nu \hbar \tilde{\rho}_1({\bf k}_1,{\bf k}_2)}
{\E_{{\bf k}_2}-\E_{{\bf k}_1}+\hbar \omega+i \nu \hbar}.
\eqno{(1.3)}
$$

The relation (1.3) represents the solution of linear
Schr\"{o}din\-ger---Boltz\-mann equation, expressed through perturbation to
equilibrium mat\-rix of density
$\tilde{\rho}_1({\bf k}_1-{\bf k}_2)=
\langle{\bf k}_1|\tilde\rho_{1}|{\bf k}_2\rangle$.
Let's find this perturbation.

Let's take advantage of the obvious relation
$$
[H-\mu,{\tilde \rho}]=0.
$$
In linear approximation from here it is received
$$
[H_0-\mu_0,{\tilde \rho}_1]+[H_1-\delta\mu,{\tilde \rho}_0]=0.
$$
Transforming the first commutator from here we receive
$$
[H_0,{\tilde \rho}_1]=[\delta\mu -H_1,{\tilde \rho}_0],
$$
or
$$
[H_0,{\tilde \rho}_1]=[\delta\mu-e\varphi,{\tilde \rho}_0].
$$

Let's designate now
$$
\delta\mu_*=\delta\mu-e\varphi.
$$

Then the previous equality write down in the form
$$
[H_0,{\tilde \rho}_1]=[\delta\mu_*,{\tilde \rho}_0].
$$

From here we receive that
$$
(\E_{{\bf k}_1}-\E_{{\bf k}_2})
\langle{\bf k}_1|{\tilde\rho_1}|{\bf k}_2\rangle
=-(f_{{\bf k}_1}-f_{{\bf k}_2})\langle{\bf k}_1|\delta\mu_*|{\bf k}_2\rangle,
$$
from which
$$
\langle{\bf k}_1|{\tilde\rho}_1|{\bf k}_2\rangle=-
\dfrac{f_{{\bf k}_1}-f_{{\bf k}_2}}
{\E_{{\bf k}_1}-\E_{{\bf k}_2}}\langle{\bf k}_1|\delta\mu_*|{\bf k}_2\rangle,
\eqno{(1.4)}
$$
or, that all the same,
$$
\tilde{\rho}_1({\bf k}_1,{\bf k}_2)=-\dfrac{f_{{\bf k}_1}-f_{{\bf k}_2}}
{\E_{{\bf k}_1}-\E_{{\bf k}_2}}\delta\mu_*({\bf k}_1-{\bf k}_2).
$$

We have received perturbation to the equilibrium matrix of the density, expressed
through perturbation of chemical potential. The last we will find from the
preservation law of numerical density.

We put next
$$
{\bf k}_1={\bf k}+\dfrac{\mathbf{k}}{2},\qquad
{\bf q}_2={\bf k}-\dfrac{\mathbf{q}}{2}
$$
and rewrite in this terms equations
(1.4), (1.5) и (1.6). We receive following equalities
$$
\hbar \omega\Big \langle {\bf k}+\dfrac{{\bf q}}{2} \Big|\rho_1\Big|{\bf k}-
\dfrac{{\bf q}}{2}\Big\rangle=
\Big(\E_{{\bf k}+{\bf q}/2}-\E_{{\bf k}-{\bf q}/2}\Big)
\Big \langle {\bf k}+\dfrac{{\bf q}}{2} \Big|\rho_1\Big|{\bf k}-
\dfrac{{\bf q}}{2}\Big\rangle-
$$\medskip
$$
-e\varphi({\bf q})(f_{{\bf k}+{\bf q}/2}-
f_{{\bf k}-{\bf q}/2})+i \nu\hbar
\Big \langle {\bf k}+\dfrac{{\bf q}}{2} \Big|\tilde \rho_1-\rho_1\Big|{\bf k}-
\dfrac{{\bf q}}{2}\Big\rangle,
\eqno{(1.2')}
$$\\
$$
 \Big\langle {\bf k}+\dfrac{{\bf q}}{2} \Big|\rho_1\Big|{\bf k}-
\dfrac{{\bf q}}{2}\Big\rangle \equiv \rho_1({\bf q})=-
\dfrac{e\varphi({\bf q})(f_{{\bf k}+{\bf q}/2}-
f_{{\bf k}-{\bf q}/2})}{\E_{{\bf k}-{\bf q}/2}-\E_{{\bf k}+{\bf q}/2}+
\hbar \omega+i \nu \hbar}+
$$ \medskip
$$
+\dfrac{i \nu \hbar  \Big\langle {\bf k}+\dfrac{{\bf q}}{2} \Big|\tilde\rho_1
\Big|{\bf k}- \dfrac{{\bf q}}{2}\Big\rangle}
{\E_{{\bf k}-{\bf q}/2}-\E_{{\bf k}+{\bf q}/2}+\hbar \omega+i \nu \hbar},
\eqno{(1.3')}
$$\medskip

$$
 \Big\langle {\bf k}+\dfrac{{\bf q}}{2} \Big|\tilde\rho_1\Big|{\bf k}-
\dfrac{{\bf q}}{2}\Big\rangle=- \dfrac{f_{{\bf k}+{\bf q}/2}-
f_{{\bf k}-{\bf q}/2})}{\E_{{\bf k}+{\bf q}/2}-\E_{{\bf k}-{\bf q}/2}}
 \Big\langle {\bf k}+\dfrac{{\bf q}}{2} \Big|\delta \mu_*\Big|{\bf k}-
\dfrac{{\bf q}}{2}\Big\rangle.
\eqno{(1.4')}
$$\medskip

\begin{center}
\bf 2.  Perturbation of chemical potential
\end{center}

The quantity $\delta\mu$ (or $\delta\mu_*$) is responsible for the local
preservation of number of particles (electrons). This local law
preservation looks like \cite{Mermin}
$$
\omega\delta n({\bf q}, \omega,\nu)=
{\bf q}\delta{\bf j}({\bf q}, \omega,\nu).
\eqno{(2.1)}
$$

In equation (2.1) $\delta n({\bf q},\omega,\nu)$,
$\delta{\bf j}({\bf q},\omega,\nu)={\bf j}({\bf q},\omega,\nu)$
are сhange of concentration and stream density of electrons under action
electric field, and
$$
\delta n({\bf q}, \omega,\nu)=\int \dfrac{d{\bf k}}{4\pi^3}
\left\langle{\bf k}+\dfrac{\mathbf{q}}{2}\Big|\rho_1\Big|{\bf k}-
\dfrac{\mathbf{q}}{2}\right\rangle,
$$
$$
\delta {\bf j}({\bf q}, \omega,\nu)=
\int \dfrac{\hbar {\bf k}d{\bf k}}{4\pi^3m}
\left\langle{\bf k}+\dfrac{\bf q}{2}\Big|\rho_1\Big|{\bf k}-\dfrac{\bf q}{2}
\right\rangle.
$$

Let's return to the equation (1.2) which have been written down
concerning perturbation to the matrix density. From this equation follows,
that
$$
\int \dfrac{\hbar \omega d{\bf k}}{4\pi^3}
\left\langle{\bf k}+
\frac{\bf q}{2}\Big|\rho_1\Big|{\bf k}-\frac{\bf q}{2}\right\rangle
=\hspace{7cm}
$$
$$
=\int \dfrac{ d{\bf k}}{4\pi^3}
\big(\E_{{\bf k}+{\bf q}/{2}}-\E_{{\bf k}-{\bf q}/{2}}\big)
\Big\langle{\bf k}+\frac{\bf q}{2}\Big|\rho_1\Big|{\bf k}-
\frac{\bf q}{2}\Big\rangle-
$$
$$
-e{\varphi}({\bf q})\int \dfrac{ d{\bf k}}{4\pi^3}
\Big(f_{{\bf k}+{\bf q}/{2}}-
f_{{\bf k}-{\bf q}/{2}}\Big)+
$$
$$\hspace{4cm}
+i\hbar\nu\int \dfrac{ d{\bf k}}{4\pi^3}
\left\langle{\bf k}+\frac{\bf q}{2}\Big|{\tilde \rho}_1-
\rho_1\Big|{\bf k}-\frac{\bf q}{2}\right\rangle.
$$

Let's transform this equality, using obvious parities
$$
\E_{{\bf k}+{\bf q}/{2}}-\E_{{\bf k}-{\bf q}/{2}}=
\dfrac{\hbar^2 {\bf k}{\bf q}}{m},
\eqno{(2.2)}
$$
and
$$
\int \dfrac{ d{\bf k}}{4\pi^3}\Big(f_{{\bf k}-{\bf q}/{2}}-
f_{{\bf k}+{\bf q}/{2}}\Big)=0.
$$

As result last equality can be rewrite in the form
$$
\hbar\int \dfrac{ d{\bf k}}{4\pi^3}\Big(\omega
\left\langle{\bf k}+\dfrac{\mathbf{q}}{2}\Big|\rho_1\Big|{\bf k}-
\dfrac{\mathbf{q}}{2}\right\rangle-\dfrac{\hbar {\bf k}{\bf q}}{m}
\left\langle{\bf k}+\dfrac{\mathbf{q}}{2}\Big|\rho_1\Big|{\bf k}-
\dfrac{\mathbf{q}}{2}\right\rangle\Big)=
$$
$$
=i\hbar\nu\int \dfrac{ d{\bf k}}{4\pi^3}
\left\langle{\bf k}+\dfrac{\mathbf{q}}{2}\Big|{\tilde \rho}_1-
\rho_1\Big|{\bf k}-\dfrac{\mathbf{q}}{2}\right\rangle.
$$

Expression in the left part of this relation according to (2.1) is equal
to zero
$$
\hbar\omega\delta n({\bf q},\omega,\nu)-
\hbar{\bf q}\delta{\bf j}({\bf q},\omega,\nu)=0.
$$

The last equality is true owing to the law of local preservation of number
particles. From here follows, that
$$
\int \dfrac{ d{\bf k}}{4\pi^3}
\left\langle{\bf k}+\dfrac{\mathbf{q}}{2}\Big|{\tilde \rho}_1-
\rho_1\Big|{\bf k}-\dfrac{\mathbf{q}}{2}\right\rangle=0.
$$
This equality is equivalent to the following
$$
\int \dfrac{ d{\bf k}}{4\pi^3}
\left\langle{\bf k}+\dfrac{\mathbf{q}}{2}\Big|{\tilde \rho}_1\Big|{\bf k}-
\dfrac{\mathbf{q}}{2}\right\rangle=\hspace{3cm}
$$
$$
\hspace{3cm}=\int \dfrac{ d{\bf k}}{4\pi^3}
\left\langle{\bf k}+\dfrac{\mathbf{q}}{2}\Big|
\rho_1\Big|{\bf k}-\dfrac{\mathbf{q}}{2}\right\rangle.
$$

Considering earlier received expression (1.4) for
$\langle{\bf k}_1|\tilde\rho_1|{\bf k}_2\rangle$
(or relation (1.4') for
$\langle {\bf k}+{\bf q}/2|\tilde \rho_1|{\bf k}-{\bf q}/2$), from here
we have
$$
\delta\mu_*({\bf q})\int \dfrac{ d{\bf k}}{4\pi^3}
\dfrac{f_{{\bf k}+{\mathbf{q}}/{2}}-f_{{\bf k}-{\mathbf{q}}/{2}}}
{\E_{{\bf k}-{\bf q}/2}-\E_{{\bf k}+{\bf q}/2}}=\hspace{3cm}
$$
$$\hspace{3cm}
=\int \dfrac{ d{\bf k}}{4\pi^3}
\langle{\bf k}+\dfrac{\mathbf{q}}{2}\Big|
\rho_1\Big|{\bf k}-\dfrac{\mathbf{q}}{2}\rangle.
$$

Thus for perturbation quantity $ \delta\mu_*({\bf q}) $ it is received
$$
\delta\mu_*({\bf q})=\dfrac{1}{B({\bf q},0)}\int
\dfrac{ d{\bf k}}{4\pi^3}\left\langle{\bf k}+\dfrac{\mathbf{q}}{2}\Big|
\rho_1\Big|{\bf k}-\dfrac{\mathbf{q}}{2}\right\rangle.
\eqno{(2.3)}
$$

Here the following designation is accepted
$$
B({\bf q},0)=
\int \dfrac{ d{\bf k}}{4\pi^3}
\dfrac{f_{{\bf k}+{\mathbf{q}}/{2}}-f_{{\bf k}-{\mathbf{q}}/{2}}}
{\E_{{\bf k}-{\bf q}/2}-\E_{{\bf k}+{\bf q}/2}}.
$$

From equation  (1.3) we obtain
$$
[\E_{{\bf k}_2}-\E_{{\bf k}_1}+\hbar \big(\omega+i\nu\big)]
\langle{\bf k}_1|\rho_1|{\bf k}_2\rangle=\hspace{3cm}
$$
$$\hspace{3cm}
=-e(f_{{\bf k}_1}-f_{{\bf k}_2}){\varphi}({\bf k}_1-{\bf k}_2)
+i\hbar\nu\langle{\bf k}_1|{\tilde \rho}_1|{\bf k}_2\rangle.
$$
Last component in this equality we will replace according to (1.4).
We receive that
$$
\Big[\E_{{\bf k}_2}-\E_{{\bf k}_1}+\hbar \big(\omega+i\nu\big)\Big]
\left\langle{\bf k}_1\left|\rho_1\right|{\bf k}_2\right\rangle=\hspace{5cm}
$$
$$\hspace{1cm}
=-e(f_{{\bf k}_1}-f_{{\bf k}_2}){\varphi}({\bf k}_1-{\bf k}_2)-
i\hbar\nu\dfrac{f_{{\bf k}_1}-f_{{\bf k}_2}}{\E_{{\bf k}_1}-\E_{{\bf k}_2}}
\delta\mu_*({\bf k}_1-{\bf k}_2).
$$

From this equation we obtain expression for
$\langle{\bf k}_1|\rho_1|{\bf k}_2\rangle$:
$$
\langle{\bf k}_1|\rho_1|{\bf k}_2\rangle=-
\dfrac{e(f_{{\bf k}_1}-f_{{\bf k}_2}){\varphi}({\bf k}_1-{\bf k}_2)}
{\hbar \omega+i\hbar\nu-\E_{{\bf k}_1}+\E_{{\bf k}_2}}-\hspace{3cm}
$$
$$\hspace{3cm}
-i\hbar\nu\dfrac{f_{{\bf k}_1}-f_{{\bf k}_2}}{\E_{{\bf k}_1}-\E_{{\bf k}_2}}
\cdot\dfrac{\delta\mu_*({\bf k}_1-{\bf k}_2)}
{\hbar \omega+i\hbar\nu-\E_{{\bf k}_1}+\E_{{\bf k}_2}},
\eqno{(2.4)}
$$
or, after decomposition on partial fractions,
$$
\langle{\bf k}_1|\rho_1|{\bf k}_2\rangle=-
\dfrac{e(f_{{\bf k}_1}-f_{{\bf k}_2}){\varphi}({\bf k}_1-{\bf k}_2)}
{\E_{{\bf k}_2}-\E_{{\bf k}_1}+\hbar \omega+i\hbar\nu}+\hspace{4cm}
$$
$$
+
\dfrac{i\nu}{\omega+i \nu}\cdot\dfrac{f_{{\bf k}_1}-f_{{\bf k}_2}}
{\E_{{\bf k}_2}-\E_{{\bf k}_1}}\delta \mu_*({\bf k}_1-{\bf k}_2)-
$$
$$-
\dfrac{i\nu}{\omega+i \nu}\cdot\dfrac{f_{{\bf k}_1}-f_{{\bf k}_2}}
{\E_{{\bf k}_2}-\E_{{\bf k}_1}+\hbar \omega+i\hbar\nu}
\delta\mu_*({\bf k}_1-{\bf k}_2),
\eqno{(2.4')}
$$

Passing to variables $ {\bf k} $ and $ {\bf q} $, from here we receive
$$
\Big\langle{\bf k}+\dfrac{{\bf q}}{2}\Big|\rho_1\Big|{\bf k}-
\dfrac{{\bf q}}{2}\Big\rangle=-
\dfrac{e(f_{{\bf k+q}/2}-f_{{\bf k-q}/2}){\varphi}({\bf q})}
{\E_{{\bf k-q}/2}-\E_{{\bf k+q}/2}+\hbar(\omega+i\nu)}+
$$
$$
+\dfrac{i\nu}{\omega+i\nu}\cdot
\dfrac{f_{{\bf k+q}/2}-f_{{\bf k-q}/2}}{\E_{{\bf k-q}/2}-\E_{{\bf k+q}/2}}
\delta\mu_*({\bf q},\omega,\nu)-
$$
$$-
\dfrac{i\nu}{\omega+i\nu}\cdot
\dfrac{(f_{{\bf k+q}/2}-f_{{\bf k-q}/2})\delta\mu_*({\bf q},\omega,\nu)}
{\E_{{\bf k-q}/2}-\E_{{\bf k+q}/2}+\hbar(\omega+i\nu)}.
\eqno{(2.4'')}
$$\medskip

Let's substitute expression (2.4") in the formula for perturbation of chemical
potential (2.3). On this way for perturbation it is received the
following expression
$$
\delta\mu_*({\bf q})=-\dfrac{(\omega+i \nu)B({\bf q},\omega+i \nu)}
{\omega B({\bf q},0)+i \nu B({\bf q},\omega+i \nu)}e{\varphi}({\bf q}).
\eqno{(2.5)}
$$

Here
$$
 B({\bf q},\omega+i \nu)=\int \dfrac{ d{\bf k}}{4\pi^3}
\dfrac{f_{{\bf k}+{\mathbf{q}}/{2}}-f_{{\bf k}-{\mathbf{q}}/{2}}}
{\E_{{\bf k}-{\bf q}/2}-\E_{{\bf k}+{\bf q}/2}+\hbar(\omega+i \nu)}.
$$

\begin{center}
\bf 3. Electric conductivity and dielectric permeability
\end{center}

Let's substitute (2.5) in (2.4") and in the received expression we will
result similar members.
As result we receive the following expression
$$
\Big \langle {\bf k}+\dfrac{{\bf q}}{2} \Big|\rho_1\Big|
{\bf k}-\dfrac{{\bf q}}{2}\Big\rangle=-e\varphi({\bf q})
\dfrac{f_{{\bf k+q}/2}-f_{{\bf k-q}/2}}
{\omega B({\bf q},0)+i \nu
B({\bf q},\omega+i \nu)} \times
$$
$$ \times
\Bigg[\dfrac{\omega B({\bf q},0)}{\E_{{\bf k-q}/2}-\E_{{\bf k+q}/2}+
\hbar(\omega+i \nu)}+\dfrac{i \nu B({\bf q},\omega+i \nu)}
{\E_{{\bf k-q}/2}-\E_{{\bf k+q}/2}}\Bigg].
\eqno{(3.1)}
$$

The current density ${\bf j}_e({\bf q}, \omega,\nu)$ is
calculated by
$\langle{\bf k}_1|\rho_1|{\bf k}_2\rangle$

$$
{\bf j}_e({\bf q}, \omega,\bar \nu)=e\,
{\bf j}({\bf q}, \omega,\bar \nu)=\dfrac{e\hbar}{m}
\int \dfrac{{\bf k}d{\bf k}}{4\pi^3}
\left\langle{\bf k}+\dfrac{\mathbf{q}}{2}\Big|\rho_1\Big|{\bf k}-
\dfrac{\mathbf{q}}{2}\right\rangle.
$$ \medskip

Thus intensity of electric field is connected with potential
of this field by relation
$
{\bf E}({\bf q}, \omega)=-i{\bf q}{\varphi}({\bf q},\omega),
$
From here the field potential is equal
$$
{\varphi}({\bf q},\omega)=
i\dfrac{{\bf q}{\bf E}({\bf q}, \omega)}{q^2}.
$$
Hence, expression for current density
${\bf j}_e({\bf q}, \omega,\bar \nu)$
by means of  relation (3.1) it is possible to rewrite in the form
$$
 {\bf j}_e({\bf q}, \omega,\nu)\equiv \sigma_l {\bf E(q},\omega)=$$$$=
-i\dfrac{e^2\hbar}{mq^2} {\bf E(q},\omega)
\int \dfrac{{\bf k\,q}d{\bf k}}{4\pi^3}
\dfrac{f_{{\bf k+q}/2}-f_{{\bf k-q}/2}}
{\omega B({\bf q},0)+i \nu
B({\bf q},\omega+i \nu)} \times
$$
$$ \times
\Bigg[\dfrac{\omega B({\bf q},0)}{\E_{{\bf k-q}/2}-\E_{{\bf k+q}/2}+
\hbar(\omega+i \nu)}+\dfrac{i \nu B({\bf q},\omega+i \nu)}
{\E_{{\bf k-q}/2}-\E_{{\bf k+q}/2}}\Bigg].
$$\medskip

From here we receive

$$
\sigma_l({\bf q},\omega, \nu)=-i\dfrac{e^2\hbar}{mq^2}
\int \dfrac{{\bf k\,q}d{\bf k}}{4\pi^3}
\dfrac{f_{{\bf k+q}/2}-f_{{\bf k-q}/2}}
{\omega B({\bf q},0)+i \nu
B({\bf q},\omega+i \nu)} \times
$$

$$ \times
\Bigg[\dfrac{\omega B({\bf q},0)}{\E_{{\bf k-q}/2}-\E_{{k+q}2}+
\hbar(\omega+i \nu)}+\dfrac{i \nu B({\bf q},\omega+i \nu)}
{\E_{{\bf k-q}/2}-\E_{{\bf k+q}/2}}\Bigg].
\eqno{(3.2)}
$$\medskip

The scalar production $\mathbf{k\,q}$ we find from relation (2.2):
$$
\mathbf{k\,q}=\dfrac{m}{\hbar^2}\Big(\E_{\mathbf{k}+\mathbf{q}/2}-
\E_{\mathbf{k}-\mathbf{q}/2}\Big).
$$

By means of this relation we will write expression for dielectric
per\-me\-abi\-lity (3.2) to the following form
$$
\sigma_l(\mathbf{q},\omega,\nu)=-i\dfrac{e^2}{q^2}
\dfrac{\omega (\omega+i \nu)B({\bf q},0)B({\bf q},\omega+i \nu)}
{\omega B({\bf q},0)+i \nu B({\bf q},\omega+i \nu)}.
\eqno{(3.3)}
$$

This formula expresses longitudinal electric conductivity into
quantum collisional plasma. On the basis of (3.3) we will write expression
for longitudinal dielectric permeability into  the quantum
collisional plasma\medskip
$$
\varepsilon_l(\mathbf{q},\omega,\nu)=1+\dfrac{4\pi e^2}{q^2}
\dfrac{(\omega+i \nu)B({\bf q},0)B({\bf q},\omega+i \nu)}
{\omega B({\bf q},0)+i \nu B({\bf q},\omega+i \nu)}.
\eqno{(3.4)}
$$\medskip

From the formula (3.4) it is visible that at $ \omega=0$ we receive
$$
\varepsilon_l(\mathbf{q},\omega,\nu)=1 +\dfrac{4\pi e^2}{q^2}B({\bf q},0).
$$
Thus, at $ \omega=0$ dielectric function does not depend
from frequency of particles collisions of plasma.

At $\nu=0$ from (3.4) we receive
$$
\varepsilon_l(\mathbf{q},\omega,\nu) =
1+\dfrac{4\pi e^2}{q^2}B({\bf q},\omega).
$$
Thus, at $ \nu=0$ dielectric function passes in
the known formula received by Klimontovich and Silin in 1952
and after that by Lindhard  in 1954.

\begin{center}
\bf 4. Comparison with Mermin' formula
\end{center}

We  will write out Mermin' formula for dielectric function
\cite{Mermin}

$$
\varepsilon_l^{\rm Mermin}({\bf q},\omega,\nu)=$$$$=1+
\dfrac{(1+i/\omega\tau)(\varepsilon^0({\bf q},
\omega+i/\tau)-1)}{1+(i/\omega\tau)(\varepsilon^0({\bf q},\omega+i/\tau)-1)/
(\varepsilon^0({\bf q},0)-1)}.
\eqno{(4.1)}
$$

In expression (4.1) $\varepsilon^0(q,0)$ is the Lindhard' dielectric
function for collisionless plasma,
$$
\varepsilon^0({\bf q},\omega)=1+\dfrac{4\pi e^2}{q^2}B({\bf q},\omega),
$$

$$
B({\bf q},\omega)=\int \dfrac{d{\bf p}}{4\pi^3}\dfrac{f_{{\bf p+q}/2}-
f_{{\bf p-q}/2}}{\E_{{\bf p-q}/2}-\E_{{\bf p+q}/2}+\omega},
$$

$$
B({\bf q},0)=\int \dfrac{d{\bf p}}{4\pi^3}\dfrac{f_{{\bf p+q}/2}-f_{{\bf p-q}/2}}
{\E_{{\bf p-q}/2}-\E_{{\bf p+q}/2}}.
$$\medskip

Let's transform the formula (4.1), noticing, that $1+i/\omega\tau = (\omega+i
\nu)/\omega $, to the form \medskip
$$
\varepsilon_l^{\rm Mermin}({\bf q},\omega,\nu)=1+\dfrac{(\omega+i \nu)
[\varepsilon^0({\bf q},\omega+i \nu)-1][\varepsilon^0({\bf q},0)-1]}
{\omega[\varepsilon^0({\bf q},0)-1]+ i \nu
[\varepsilon^0({\bf q},\omega+i \nu)-1]}.
\eqno{(4.2)}
$$
We rewrite Mermin's formula (4.2) in terms of integrals $B({\bf q},\omega)$
and $B({\bf q},0)$
$$
\varepsilon_l^{\rm Mermin}({\bf q},\omega,\nu)=1+\dfrac{4\pi e^2}{q^2}
\dfrac{(\omega+i \nu)B({\bf q},\omega+i \nu)B({\bf q},0)}
{\omega B({\bf q},0)+i \nu B({\bf q},\omega+i \nu)}.
\eqno{(4.3)}
$$

Comparison of Mermin' formula  (4.3) with the formula (3.4) shows their full
сoincidence.

\begin{center}\bf
5. Quantum degenerate plasma
\end{center}

Let's calculate integrals
$$
B({\bf q},\omega+i \nu)=\int \dfrac{d{\bf k}}{4\pi^3}\dfrac{f_{{\bf k+q}/2}-
f_{{\bf k-q}/2}}{\E_{{\bf k-q}/2}-\E_{{\bf k+q}/2}+\hbar(\omega+ i \nu)}.
\eqno{(5.1)}
$$

For degenerate plasma in formula (5.1) there are designations
$$
f_{{\bf k}}=\Theta(E_F-E_{{\bf k}}),\qquad
\E_{{\bf k}}=\dfrac{\hbar^2{\bf k}^2}{2m},\qquad \E_F=\dfrac{p_F^2}{2m},
$$
$\Theta(x)$ is the Heaviside function,
$$
\Theta(x)=\left\{\begin{array}{c}
                   1,\qquad x>0, \\
                   0,\qquad x<0.
                 \end{array}\right.
$$

Let's present integral (5.1) in the form of a difference of two integrals.
We will enter in these integrals obvious replacement of variables.
It is as a result received
$$
B({\bf q},\omega+i \nu)=
$$
$$
=\int \dfrac{d{\bf k}}{4\pi^3}\dfrac{f_{{\bf k}}
[2\E_{\bf k}-\E_{\bf k+q}-\E_{\bf k-q}]}{[\E_{{\bf k-q}}-\E_{\bf k}+
\hbar(\omega+ i \nu)][\E_{\bf k}-\E_{\bf k+q}+\hbar(\omega+i \omega)]}.
\eqno{(5.2)}
$$\medskip

Let's pass to integration on the dimensionless wave vector  ${\bf P}=
\dfrac{{\bf k}}{k_F}$, vectors ${\bf P}$ and ${\bf q}$
let's direct lengthways axes $x $, believing
${\bf P}=P_x(1,0,0)$ and $ {\bf q}=q(1,0,0)$. We introduce
one more wave vector ${\bf k}=\dfrac{{\bf q}}{k_F}$. Then
$$
\E_{{\bf k-q}}-\E_{\bf k}+\hbar(\omega+ i \nu)=-\dfrac{\hbar^2 (2{\bf k\,q}+
q^2)}{2m}+\hbar(\omega+i \nu)=
$$

$$=-\dfrac{2\hbar^2q}{2m}(k_x-\dfrac{q}{2})+
\hbar(\omega+i \nu)=-2k\E_F(P_x-\dfrac{k}{2})+\hbar(\omega+i \nu)=
$$

$$
=-2k\E_F\Big(P_x-\dfrac{k}{2}-\dfrac{\hbar(\omega+i \nu)}{2k\E_F}\Big)=
-2k\E_F\Big(P_x-\dfrac{k}{2}-\dfrac{z}{k}\Big).
$$

Here
$$
z=\dfrac{\hbar(\omega+i \nu)}{2E_F}=\dfrac{\omega+i \nu}{k_Fv_F},\qquad
x=\dfrac{\omega}{k_Fv_F},\qquad y=\dfrac{\nu}{k_Fv_F}.
$$

Similarly,
$$
\E_{{\bf k}}-\E_{\bf k+q}+\hbar(\omega+ i \nu)=
-2k\E_F\Big(P_x+\dfrac{k}{2}-\dfrac{z}{k}\Big),
$$
the kernel of integral (5.2) is equal:
$$
\dfrac{2\E_{\bf k}-\E_{\bf k+q}-\E_{\bf k-q}}{[\E_{{\bf k-q}}-\E_{\bf k}+
\hbar(\omega+ i \nu)][\E_{\bf k}-\E_{\bf k+q}+\hbar(\omega+i \omega)]}=
$$

$$=-
\dfrac{1}{2\E_F\Big[\Big(P_x-\dfrac{z}{k}\Big)^2-\Big(\dfrac{k}{2}\Big)^2\Big]}.
$$

In the integral (5.2)
$$
f_{\bf k}=\Theta(\E_F-\E_F {\bf P}^2)=\Theta(1-{\bf P}^2), \qquad
\E_{\bf k}=\E_F {\bf P}^2.
$$

Now the integral (5.2) is equal
$$
B(k,z)=-\dfrac{k_F^3}{8\pi^3 \E_F}\int \dfrac{\Theta(1-{\bf P}^2)}
{(P_x-z/k)^2-(k/2)^2}=
$$
$$
=-\dfrac{3N}{4\pi mv_F^2} \int\limits_{{\bf P}^2<1}\dfrac{d^3P}
{(P_x-z/k)^2-(k/2)^2}.
$$

It is easy to see, that
$$
B(k,z)=-\dfrac{3N}{4mv_F^2}\int\limits_{-1}^{1}\dfrac{(1-P_x^2)dP_x}
{(P_x-z/k)^2-(k/2)^2}=
$$
$$
=-\dfrac{3N}{4mv_F^2}\Bigg[-2+\dfrac{1}{k}\Big[1-\Big(\dfrac{z}{k}\Big)^2-
\Big(\dfrac{k}{2}\Big)^2\Big]
\ln\dfrac{(z/k)^2-1+k-(k/2)^2}{(z/k)^2-1-k-(k/2)^2}-
$$
$$
-\dfrac{z}{k}\ln\dfrac{(1-z/k)^2-(k/2)^2}{(1+z/k)^2-(k/2)^2}
\Bigg].
$$

From this equality follows, that
$$
B(k,0)=-\dfrac{3N}{4mv_F^2}\int\limits_{-1}^{1}\dfrac{(1-P_x^2)dP_x}
{P_x^2-(k/2)^2}=\hspace{4cm}
$$
$$
\hspace{3cm}=-\dfrac{3N}{4mv_F^2}\bigg[-2+\dfrac{(2-k)(2+k)}{2k}
\ln\dfrac{k-2}{k+2}\bigg].
$$

By means of these integrals it is found electric conductivity
$$
\sigma_l(x,y,k)=i\dfrac{3e^2N \omega}{4mv_Fk_F^2k^2}\dfrac{(x+iy)b(k,z)b(k,0)}
{xb(k,0)+iyb(k,z)},
\eqno{(5.3)}
$$
where
$$
b(k,z)=\int\limits_{-1}^{1}\dfrac{(1-\tau^2)d\tau}{\tau^2-(k/2)^2}=
$$
$$
=-2+\dfrac{1}{k}\Big[1-(z/k)^2-(k/2)^2\Big]
\ln\dfrac{(z/k)^2-1+k-(k/2)^2}{(z/k)^2-1-k-(k/2)^2}-
$$
$$
-\dfrac{z}{k}\ln\dfrac{(1-z/k)^2-(k/2)^2}{(1+z/k)^2-(k/2)^2},
$$
$$
b(k,0)=\int\limits_{-1}^{1}\dfrac{(1-\tau^2)d\tau}{\tau^2-(k/2)^2}=
-2+\dfrac{(2-k)(2+k)}{2k}
\ln\dfrac{k-2}{k+2}.
$$

In the formula (5.3) we will allocate static conductivity
$\sigma_0=\dfrac{e^2N}{m\nu}$. Then
$$
\dfrac{\sigma_l(x,y,k)}{\sigma_0}=
i\dfrac{3}{4}\cdot\dfrac{\hbar \omega}{mv_F^2}\cdot
\dfrac{\hbar \nu}{mv_F^2}\cdot\dfrac{1}{k^2}\cdot\dfrac{(x+iy)b(k,0)b(k,z)}
{xb(k,0)+iy b(k,z)}.
\eqno{(5.4)}
$$

By means of (5.4) we will write the formula for dielectric permeability
$$
\varepsilon_l(x,y,k)=1-\dfrac{3\omega_p^2}{4\omega^2}\cdot
\Big(\dfrac{\hbar \omega}{mv_F^2}\Big)^2\cdot\dfrac{1}{k^2}\cdot
\dfrac{(x+iy)b(k,z)b(k,0)}{xb(k,0)+iy b(k,z)},
\eqno{(5.5)}
$$
or, in equivalent form
$$
\varepsilon_l(x,y,k)=1-\dfrac{3}{4k^2}\cdot
\Big(\dfrac{\omega_p\hbar}{mv_F^2}\Big)^2\cdot
\dfrac{(x+iy)b(k,z)b(k,0)}{xb(k,0)+iy b(k,z)}.
\eqno{(5.5')}
$$
Having entered dimensionless plasma (Langmuir) frequency
$$
x_p=\dfrac{\omega_p\hbar}{mv_F^2}=\dfrac{\omega_p\hbar}{p_Fv_F}=
\dfrac{\omega_p}{k_Fv_F},
$$
let's copy the formula (5.5 ') in the form
$$
\varepsilon_l(x,y,k)=1-\dfrac{3x_p^2}{4k^2}\cdot
\dfrac{(x+iy)b(k,z)b(k,0)}{xb(k,0)+iy b(k,z)}.
\eqno{(5.5'')}
$$ \medskip

\begin{center}
\bf   6. Degenerate classical (Fermi) plasma
\end{center}

We take expression of dielectric function for classical
degenerate plas\-mas
$$
\varepsilon_l^{\rm classic}(\omega,\nu,q)=1+\dfrac{3\omega_p^2}{q^2v_F^2}\cdot
\dfrac{1-(1-i\omega\tau)T_0(\omega,\nu,q)}{1-T_0(\omega,\nu,q)},
\eqno{(6.1)}
$$
where $q$ is the wave number,
$$
T_0(\omega,\nu,q)=\dfrac{1}{2}\int\limits_{-1}^{1}\dfrac{d \mu}{1-i \omega\tau
+iqv_F\tau \mu}.
$$

Let's result the formula (6.1) to the obvious form
$$
\varepsilon_l^{\rm classic}(\omega,\nu,q)=1+\dfrac{3\omega_p^2}{q^2v_F^2}\cdot
\dfrac{1+\dfrac{\omega+i \nu}{2qv_F}\ln\dfrac{(\omega+i \nu)/v_F-q}
{(\omega+i \nu)/v_F+q}}{1+\dfrac{i \nu}{2qv_F}\ln\dfrac{(\omega+i \nu)/v_F-q}
{(\omega+i \nu)/v_F+q}}.
\eqno{(6.2)}
$$
Let's copy the formula (6.2) in dimensionless variables
$$
x=\dfrac{\omega}{k_Fv_F},\qquad y=\dfrac{\nu}{k_Fv_F},\qquad k=\dfrac{q}{k_F},
\qquad x_p=\dfrac{\omega_p}{k_Fv_F}.
$$
We obtain that
$$
\varepsilon_l^{\rm classic}(x,y,k)=
1+\dfrac{3x_p^2}{k^2}\cdot\dfrac{1+\dfrac{z}{2k}
\ln\dfrac{z-k}{z+k}}{1+\dfrac{iy}{2k}\ln\dfrac{z-k}{z+k}},
\eqno{(6.3)}
$$
where $z=x+iy$.

In some questions instead of the formula (6.3) it is more convenient to use
following expression of dielectric function
$$
\varepsilon_l^{\rm classic}(x,y,k)=1+\dfrac{3x_p^2}{k^2}\cdot\dfrac{1+zb_0(z,k)}
{1+iyb_0(z,k)},
\eqno{(6.4)}
$$
where
$$
b_0(z,k)=\dfrac{1}{2k}\int\limits_{-1}^{1}\dfrac{d\mu}{\mu-z/k}
=\dfrac{1}{2k}\ln\dfrac{z-k}{z+k}.
$$

In all drawings dimensionless plasma frequency
is taken to equal unit: $x_p=1$.

On figs. 1--8 graphics of the real and imaginary parts
of dielectric function of quantum collisional degenerate plasmas are given.
On figs. 1 and 2 curves $1,2,3$ correspond to functions $\varepsilon_l(k)=$
$\varepsilon_l(0.1,0.001,k)$,  $\varepsilon_l(k)=
\varepsilon_l(0.15,0.001,k)$, $\varepsilon_l(k)=
\varepsilon_l(0.2,0.001,k)$.

On figs. 3 and 4 curves $1,2,3$ correspond to functions
$\varepsilon_l(k)=$\\
$\varepsilon_l(0.9,0.001,k)$, $\varepsilon_l(k)=
\varepsilon_l(1.0,0.001,k)$, $\varepsilon_l(k)=
\varepsilon_l(1.1,0.001,k)$.

On figs. 5 and 6 curves $1,2,3$ correspond to functions $\varepsilon_l(x)=$\\$
\varepsilon_l(x,0.001,0.1)$, $\varepsilon_l(x)=
\varepsilon_l(x,0.001,0.12)$, $\varepsilon_l(x)=
\varepsilon_l(x,0.001,0.15)$.

On figs. 7 and 8 curves $1,2,3$ correspond to functions $\varepsilon_l(y)=$\\$
\varepsilon_l(0.11,0.001,y)$, $\varepsilon_l(y)=
\varepsilon_l(0.12,0.001,y)$, $\varepsilon_l(y)=
\varepsilon_l(0.13,0.001,y)$.

On figs. 9--12 comparison of the real and imaginary parts of
dielectric function quantum and classical collisional degenerate
plasmas at $y=0.001$ is given.
On figs. 9 and 10 the case $x=0.3$ is represented,
on figs. 11 and 12 the case $x=1$ is represented. Curves 1 and 2
correspond to
to quantum $\varepsilon_l$ and classical $\varepsilon_l^{\rm classic}$
plasmas.

On figs. 13 and 14 the module of derivative $|d\varepsilon_l/dk|$
of dielectric function of quantum (curves 1) and classical (curves 2)
plasmas for the case $x=0.5$ at $y=0.0001$ (fig. 13) and $y=0.001$ (fig.
14) is represented.

\begin{center}
\bf 7. Conclusion
\end{center}

In the present work the detailed conclusion of formulas for the
longi\-tu\-di\-nal electric conductivity and dielectric permeability
of quantum degenerate collisional  plasmas is resulted.

For this purpose the kinetic the equation with integral of collisions
in relaxation form in momentum space is used.

Graphic research of properties the real and imaginary parts of
dielect\-ric function of quantum collisional plasmas and graphic
comparison of the real and imaginary parts of dielectric function
between quantum and classical plasmas is shown.

\begin{figure}[ht]\center
\includegraphics[width=16.0cm, height=10cm]{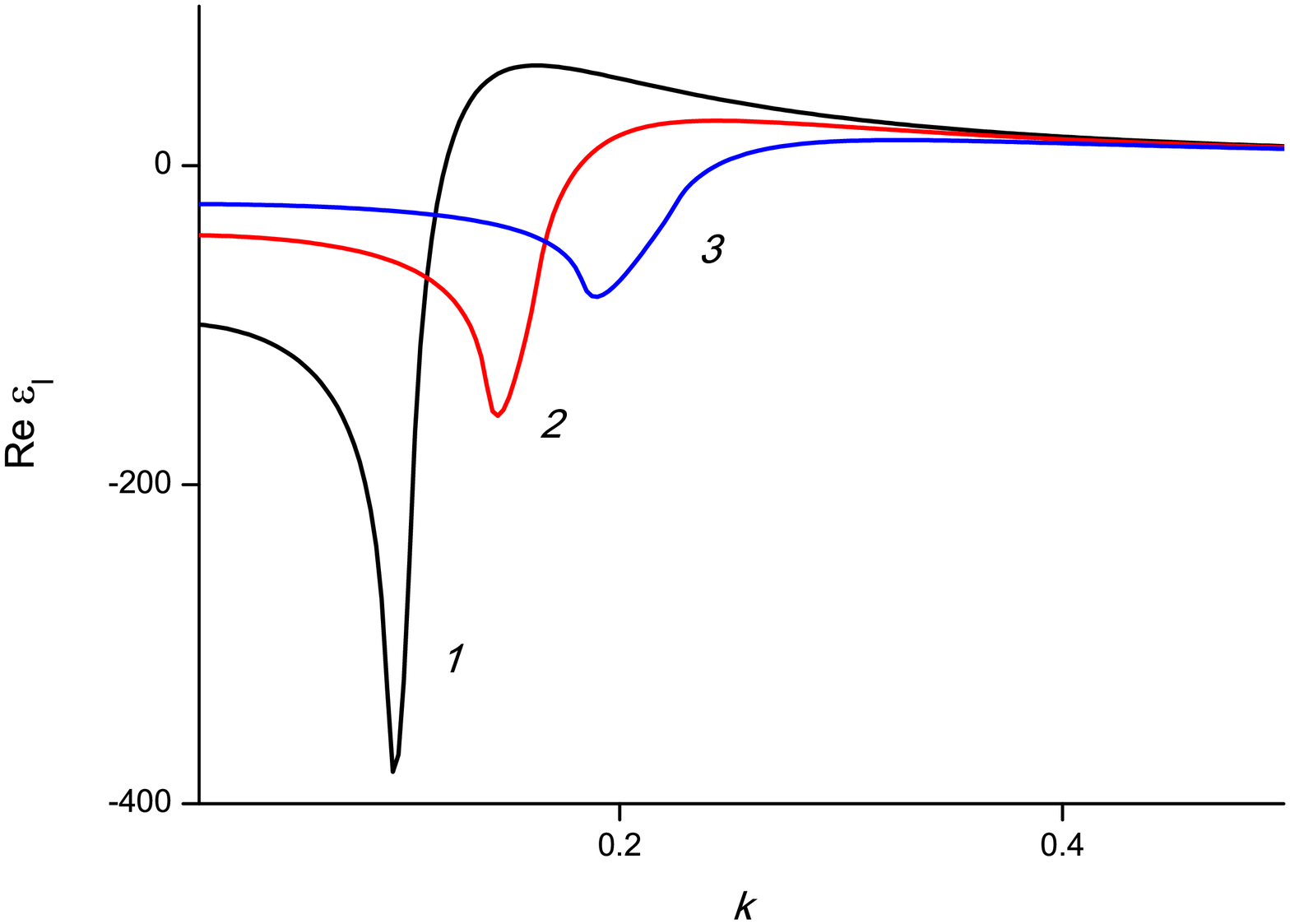}
\center{Fig. 1. Real parts of dielectric function, $x_p=1$,
$y=0.001$. Curves $1,2,3$ correspond to values of dimensionless frequency
of electric field $x=0.1, 0.15, 0.2$.}
\includegraphics[width=17.0cm, height=10cm]{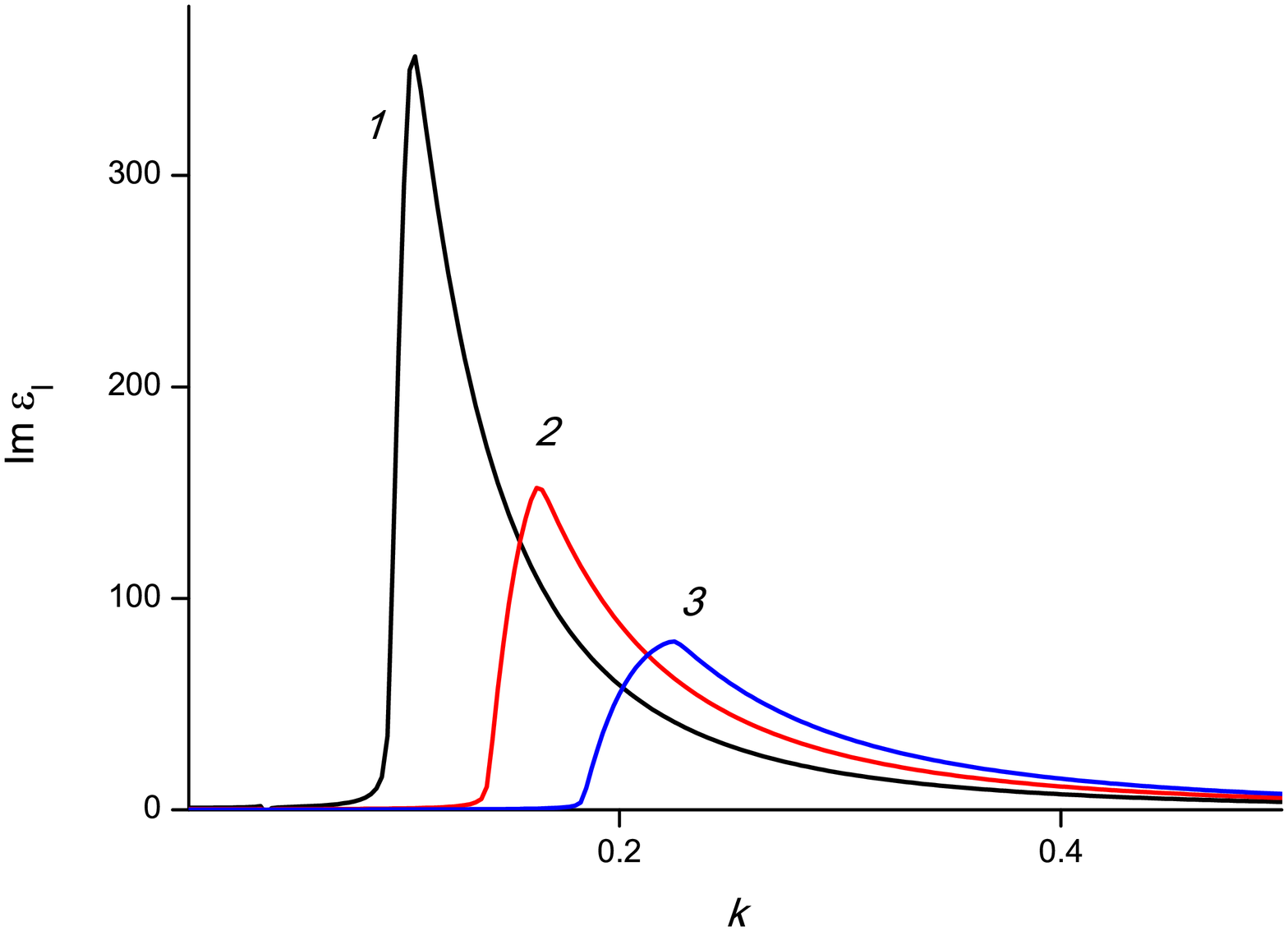}
\center{Fig. 2. Imaginary parts of dielectric function, $x_p=1$,
$y=0.001$. Curves $1,2,3$ correspond to values of dimensionless frequency
of electric field  $x=0.1, 0.15, 0.2$.}
\end{figure}

\begin{figure}[ht]\center
\includegraphics[width=16.0cm, height=10cm]{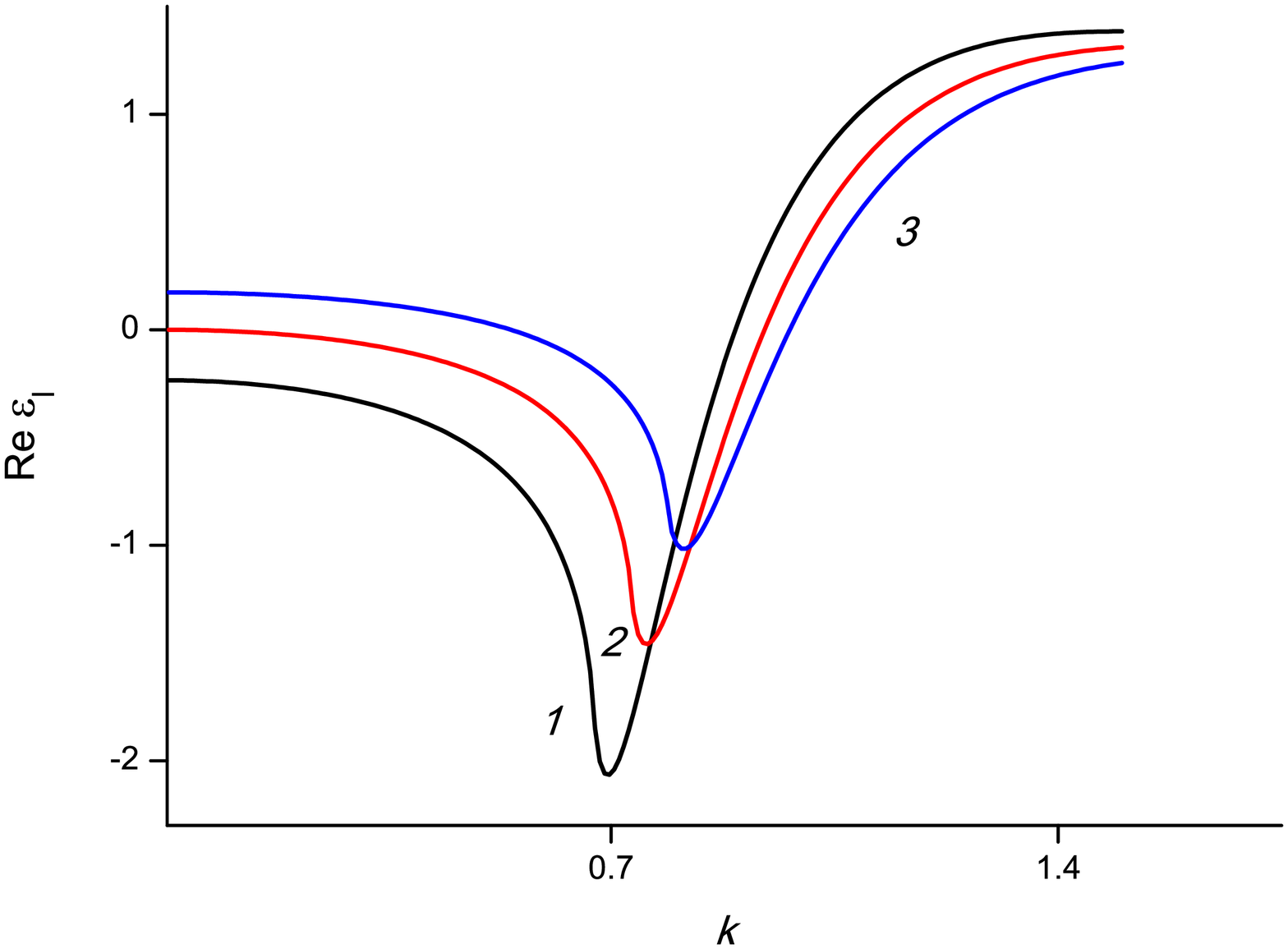}
\center{Fig. 3.  Real parts of dielectric function, $x_p=1$,
$y=0.001$. Curves $1,2,3$ correspond to values of dimensionless frequency
of electric field $x=0.9, 1.0, 1.1$.}
\includegraphics[width=17.0cm, height=10cm]{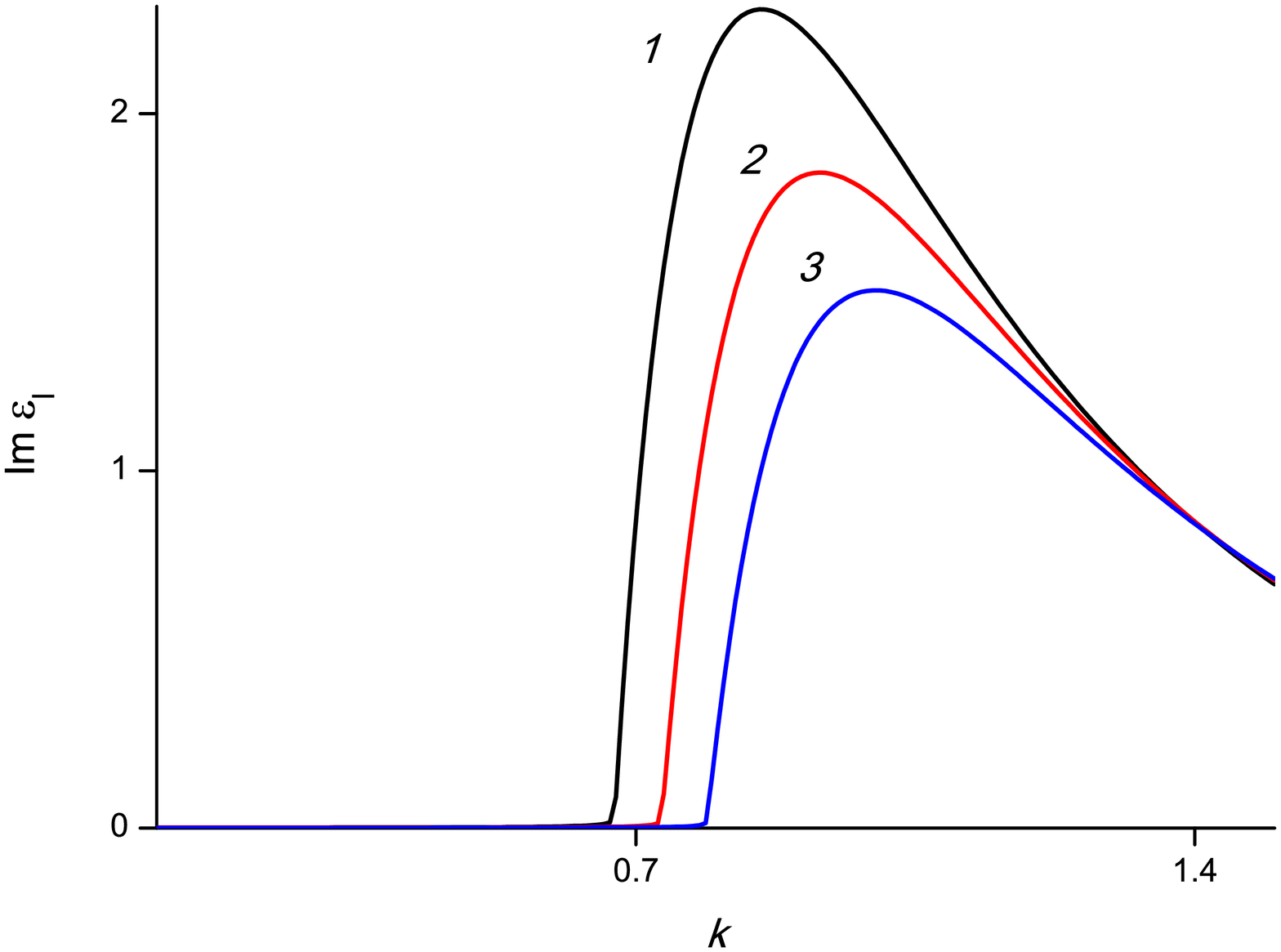}
\center{Fig. 4. Imaginary parts of dielectric function, $x_p=1$,
$y=0.001$. Curves $1,2,3$ correspond to values of dimensionless frequency
of electric field $x=0.9, 1.0, 1.1$.}
\end{figure}

\begin{figure}[ht]\center
\includegraphics[width=16.0cm, height=10cm]{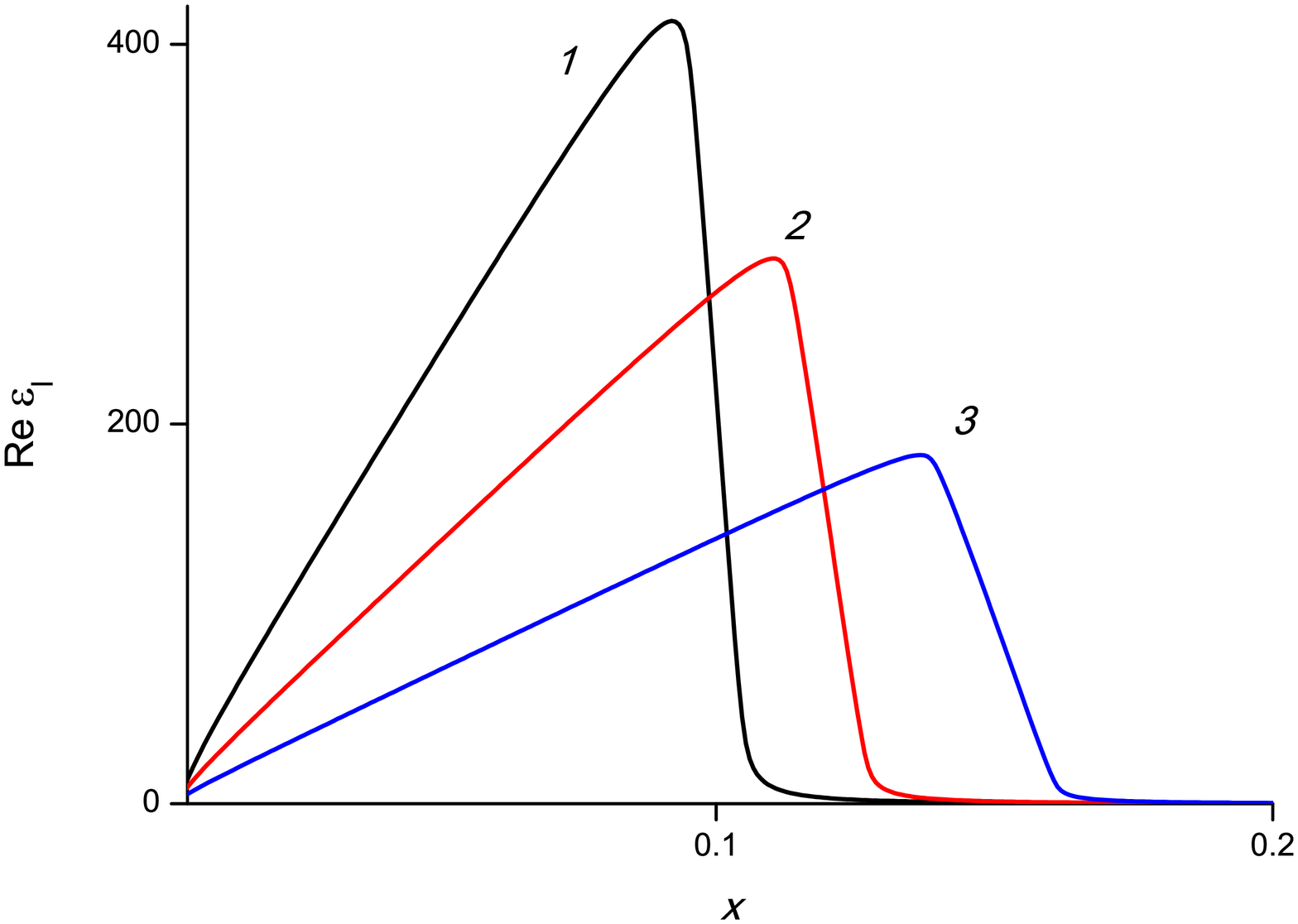}
\center{Fig. 5.  Real parts of dielectric function, $x_p=1$,
$y=0.001$. Curves $1,2,3$ correspond to values of dimensionless wave number
$k=0.1, 0.12, 0.15$.}
\includegraphics[width=17.0cm, height=10cm]{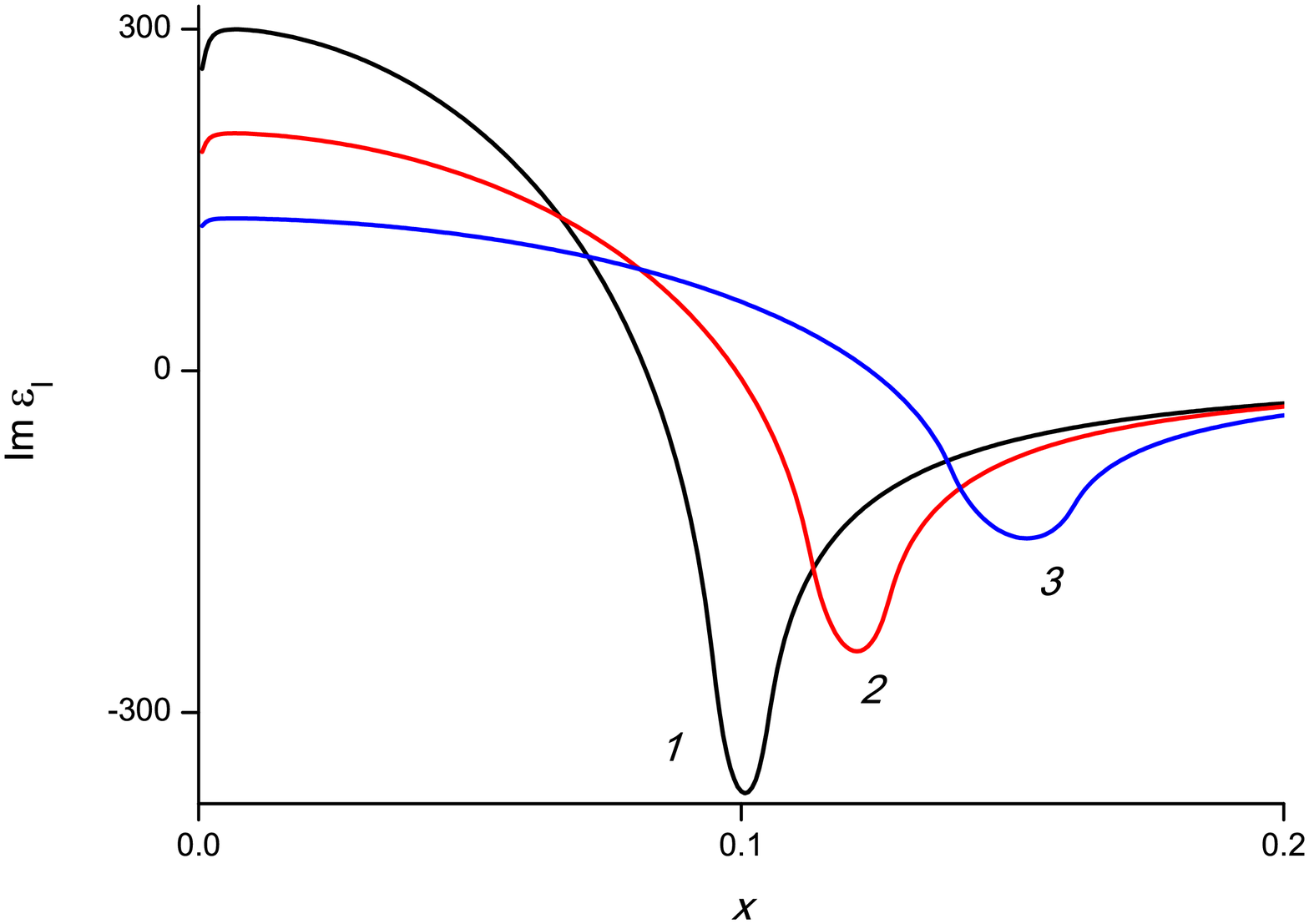}
\center{Fig. 6. Imaginary parts of dielectric function, $x_p=1$,
$y=0.001$. Curves $1,2,3$ correspond to values of dimensionless wave number
$k=0.1, 0.12, 0.15$.}
\end{figure}

\begin{figure}[ht]\center
\includegraphics[width=16.0cm, height=10cm]{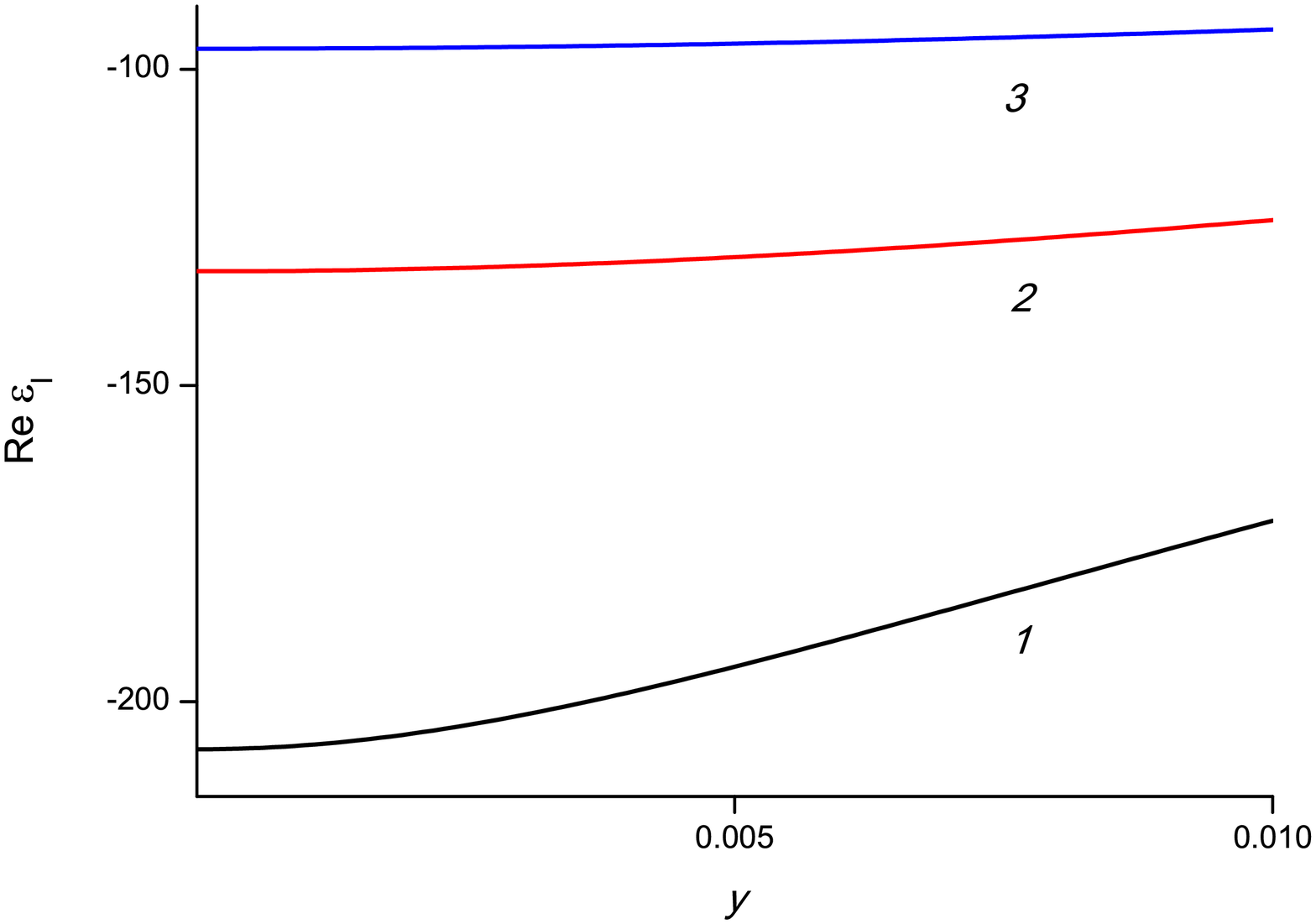}
\center{Fig. 7. Real parts of dielectric function, $x_p=1$,
$k=0.1$. Curves $1,2,3$ correspond to values of dimensionless collision
frequency $x=0.11, 0.12, 0.13$.}
\includegraphics[width=17.0cm, height=10cm]{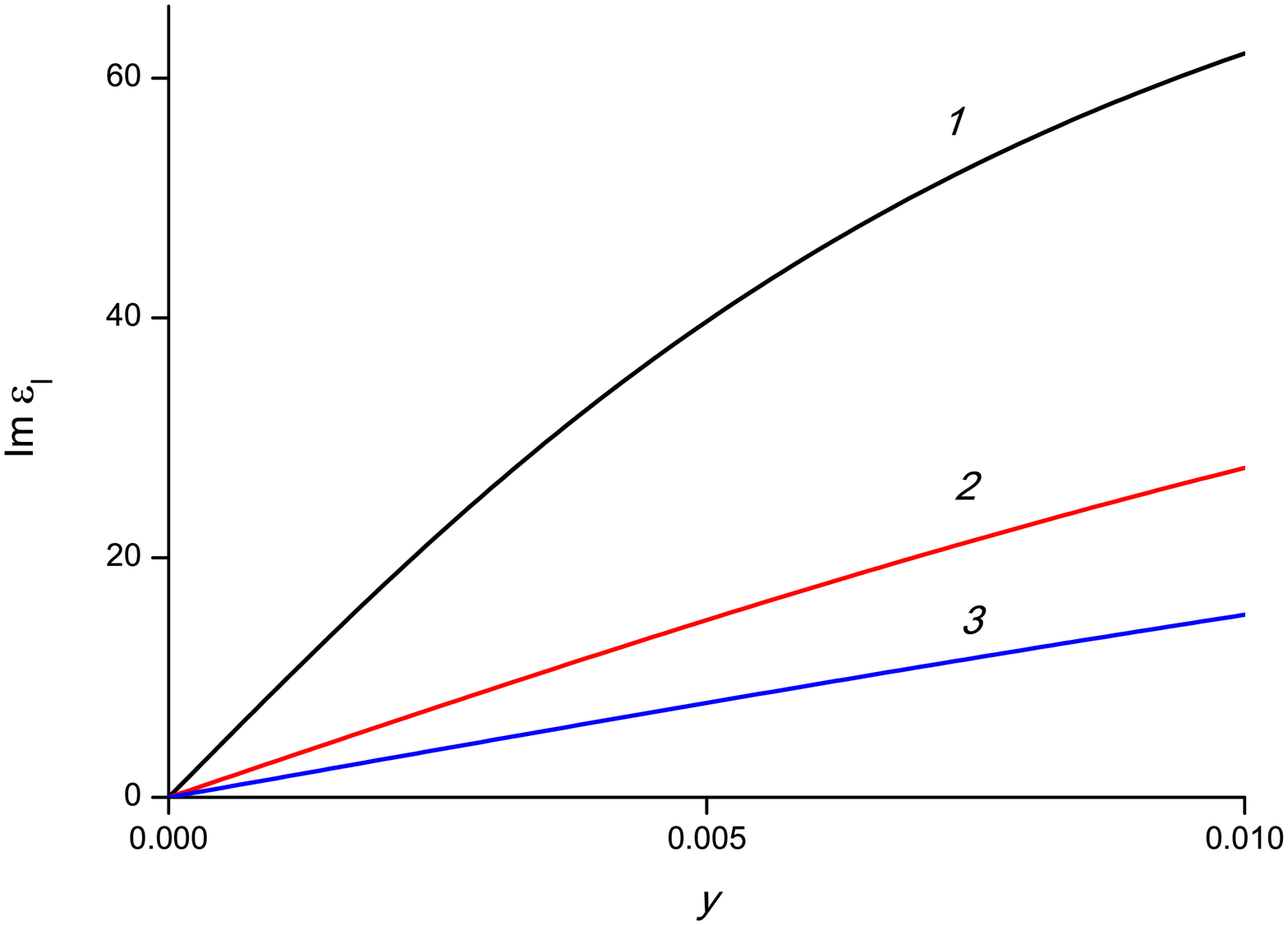}
\center{Fig. 8. Imaginary parts of dielectric function, $x_p=1$,
$k=0.1$. Curves $1,2,3$ correspond to values of dimensionless collision
frequency $x=0.11, 0.12, 0.13$.}
\end{figure}

\begin{figure}[ht]\center
\includegraphics[width=16.0cm, height=10cm]{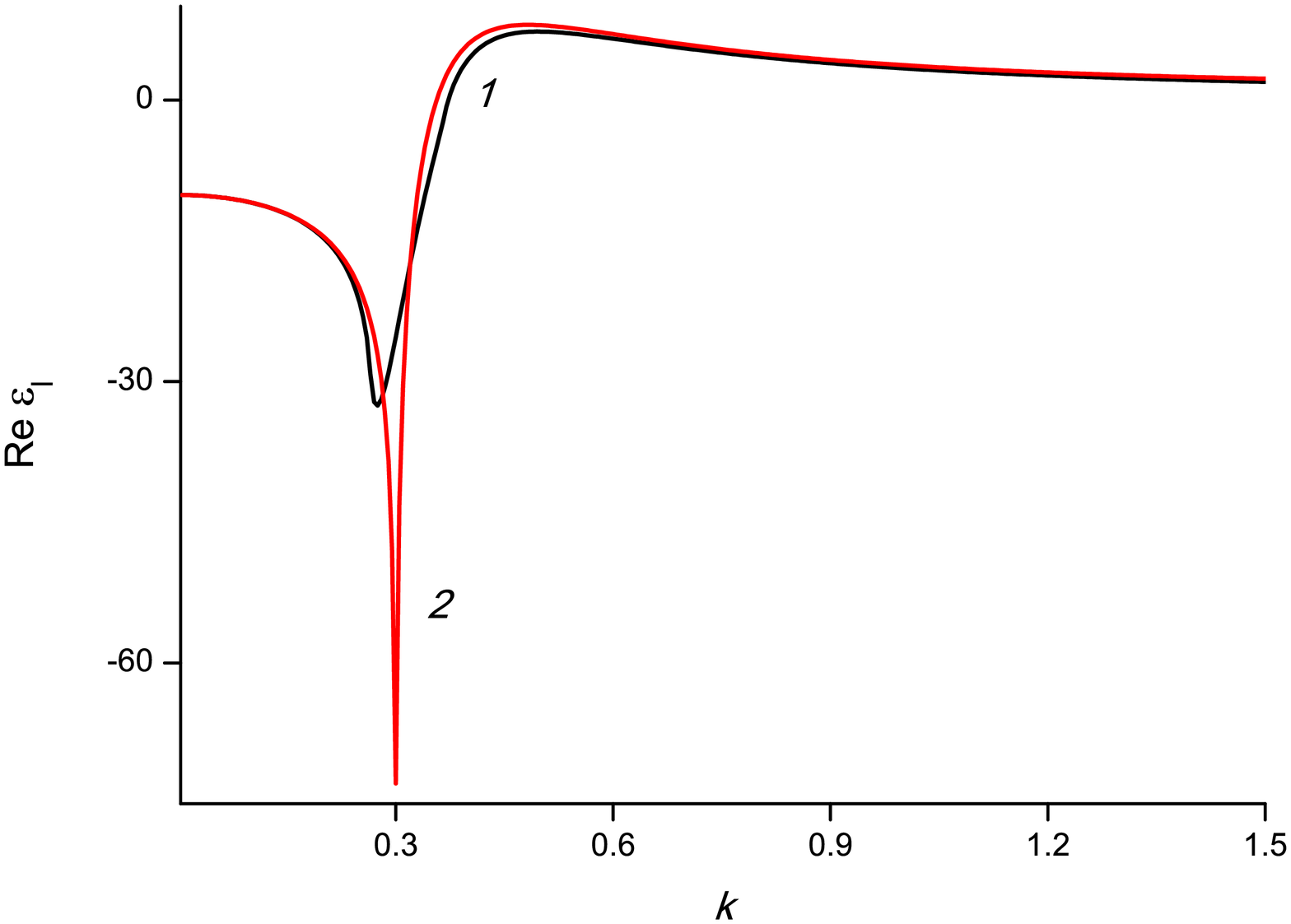}
\center{Fig. 9. Real parts of dielectric function, $x_p=1$, $x=0.3$,
$y=0.001$. Curves $1$ and $2$ correspond to quantum and classical plasmas.}
\includegraphics[width=17.0cm, height=10cm]{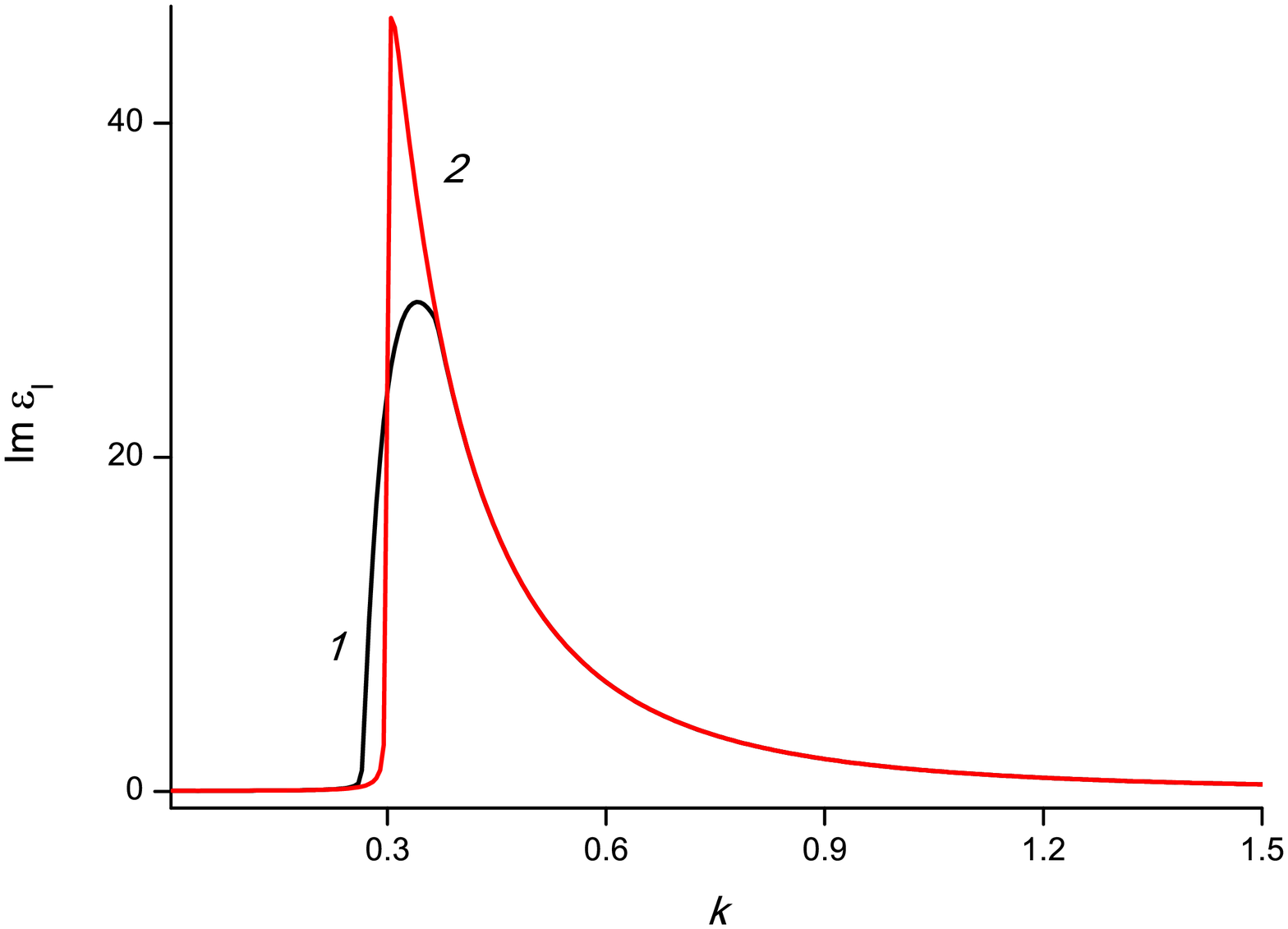}
\center{Fig. 10. Imaginary parts of dielectric function, $x_p=1$, $x=0.3$,
$y=0.001$. Curves $1$ and $2$ correspond to quantum and classical plasmas.}
\end{figure}

\begin{figure}[ht]\center
\includegraphics[width=16.0cm, height=10cm]{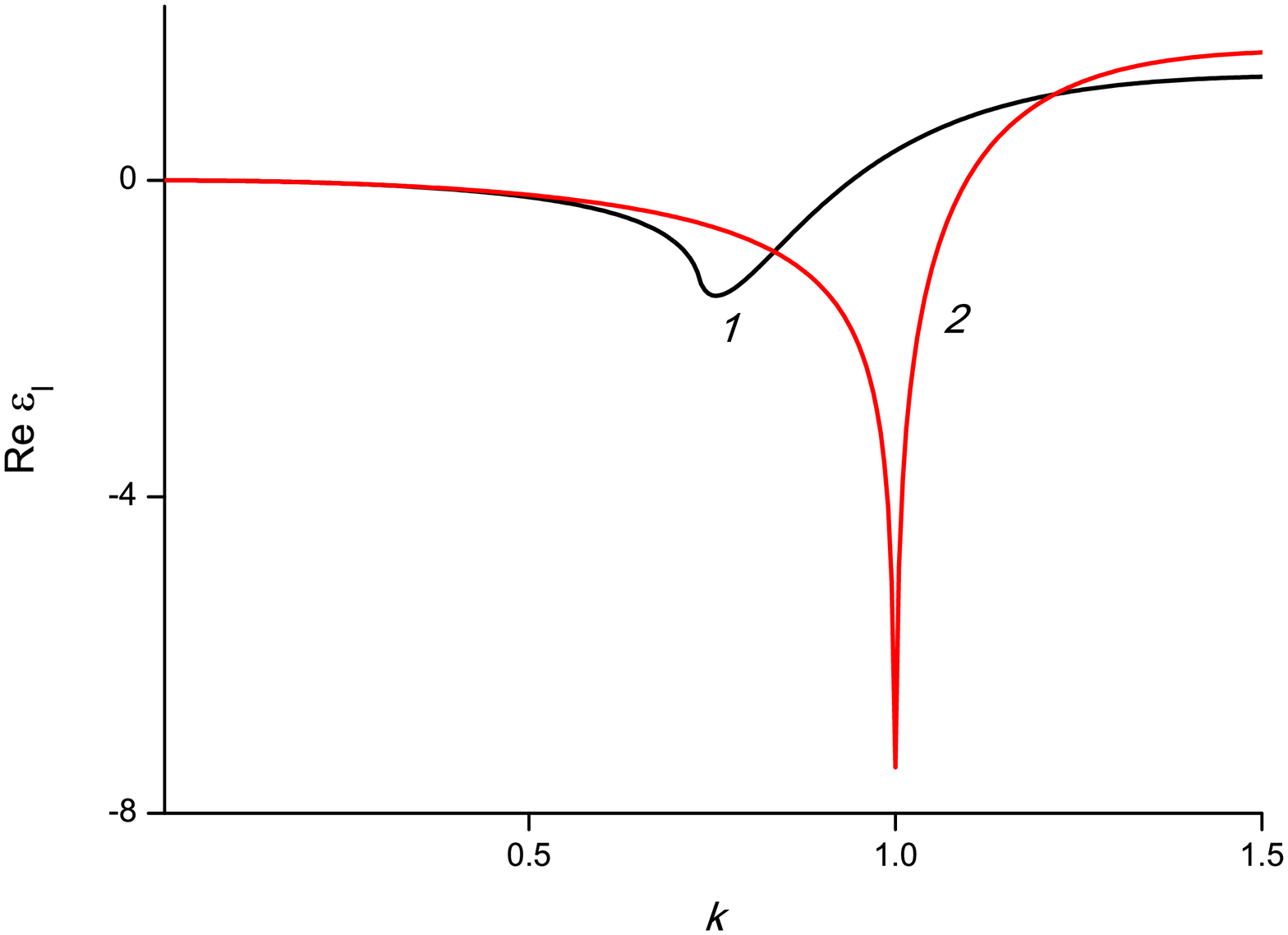}
\center{Fig. 11. Real parts of dielectric function, $x_p=1$, $x=1$,
$y=0.001$.  Curves $1$ and $2$ correspond to quantum and classical plasmas.}
\includegraphics[width=17.0cm, height=10cm]{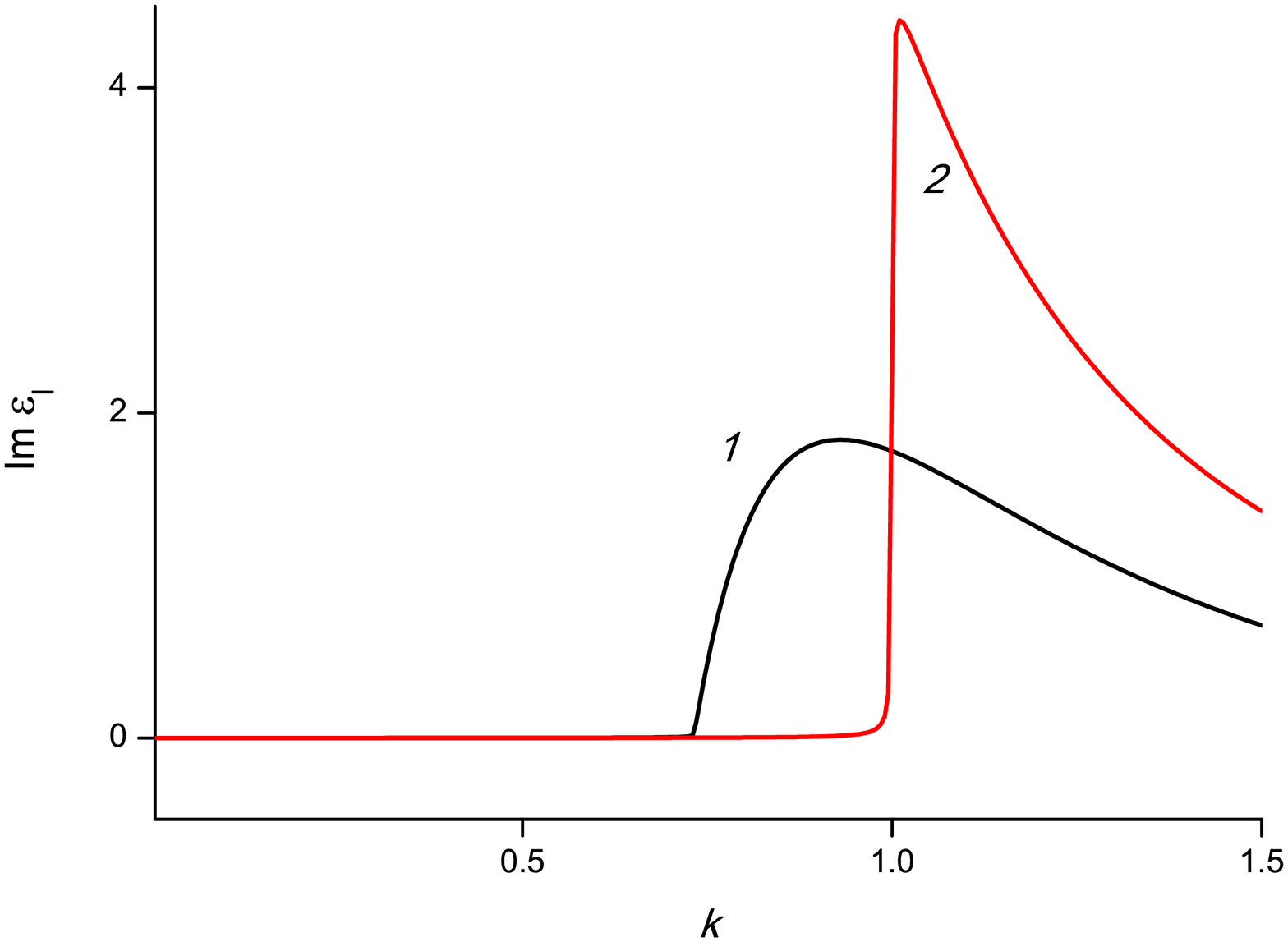}
\center{Fig. 12. Imaginary parts of dielectric function, $x_p=1$, $x=1$,
$y=0.001$.  Curves $1$ and $2$ correspond to quantum and classical plasmas.}
\end{figure}

\begin{figure}[ht]\center
\includegraphics[width=16.0cm, height=10cm]{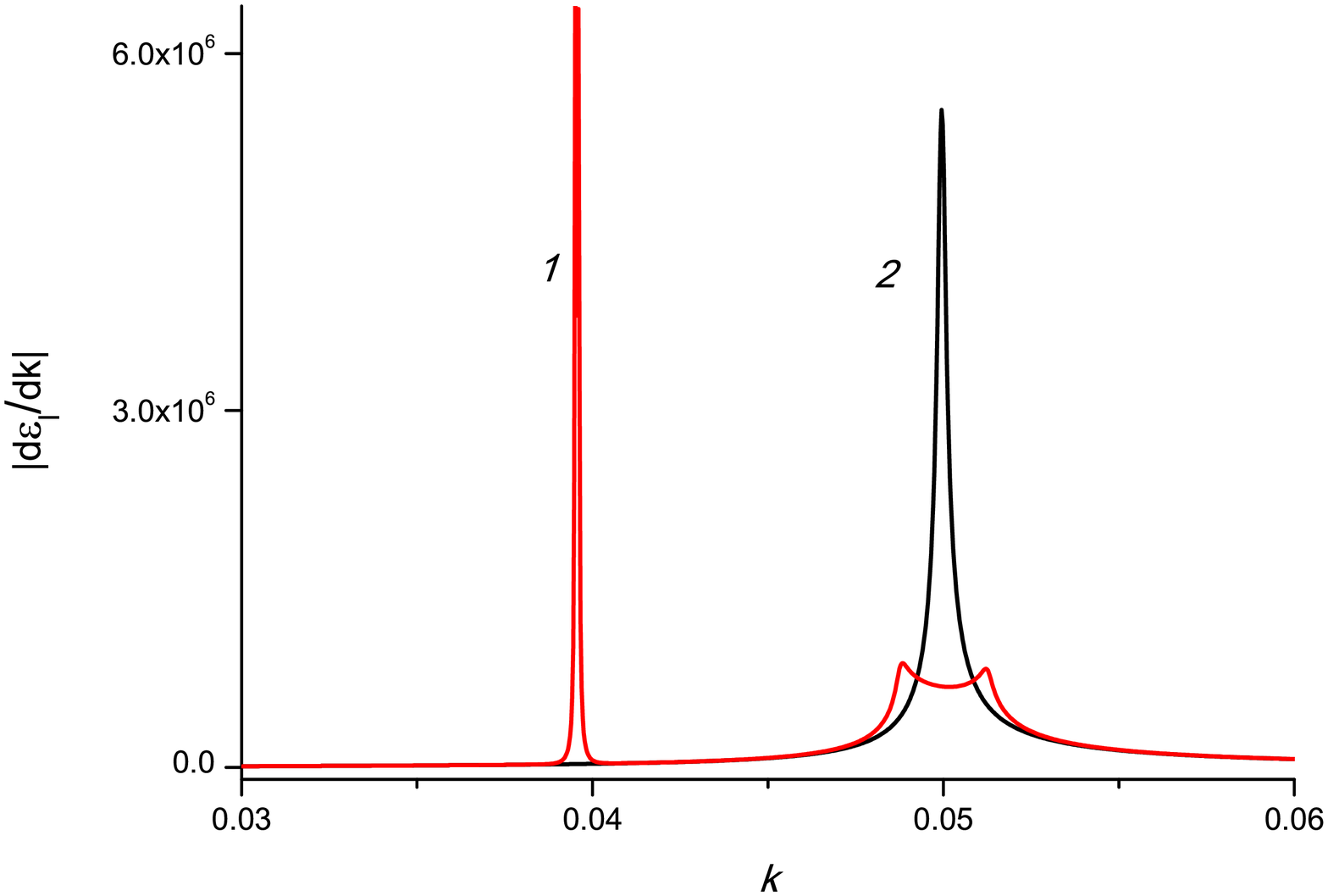}
\center{Fig. 13. Module of derivative $|d\varepsilon_l/dk|$ of dielectric
function, $x_p=1$, $x=0.05$,
$y=0.0001$. Curves $1$ and $2$ correspond to quantum and classical plasmas.}
\includegraphics[width=17.0cm, height=10cm]{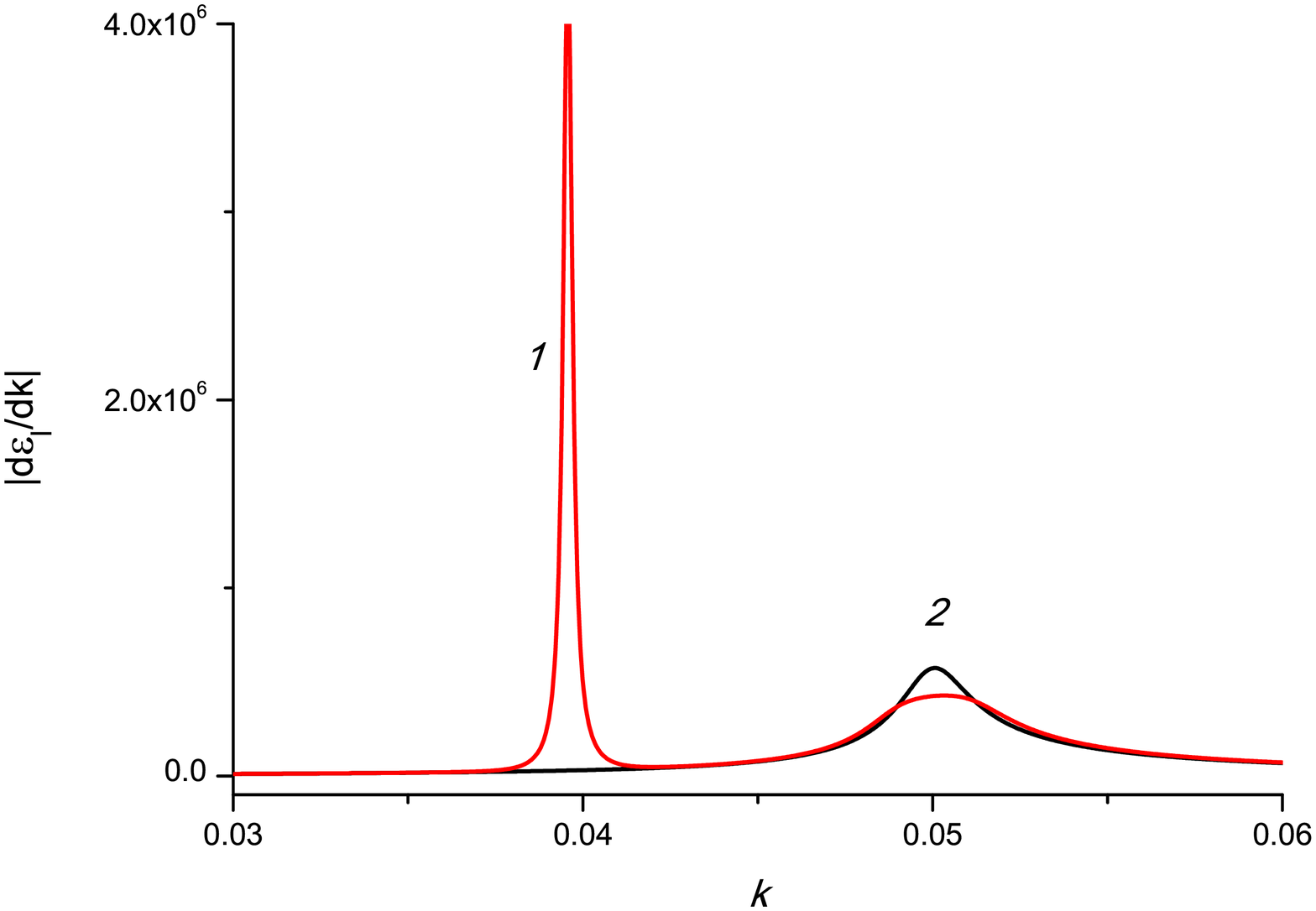}
\center{Fig. 14. Module of derivative $|d\varepsilon_l/dk|$ of dielectric
function, $x_p=1$, $x=0.05$,
$y=0.001$. Curves $1$ and $2$ correspond to quantum and classical plasmas.}
\end{figure}

\end{document}